# Versatile direct-writing of dopants in a solid state host through recoil implantation

Johannes E. Fröch,[1] Alan Bahm,[2] Mehran Kianinia,[1] Mu Zhao,[3] Vijay Bhatia,[4] Sejeong Kim,[1] Julie M. Cairney,[4] Weibo Gao,[3] Carlo Bradac,[1,5] Igor Aharonovich[1,6] and Milos Toth[1,6]

[1] School of Mathematical and Physical Sciences, University of Technology Sydney, Ultimo, New South Wales 2007, Australia

[2] Thermo Fisher Scientific, Hillsboro, Oregon 97124, USA

[3] Division of Physics and Applied Physics, School of Physical and Mathematical Sciences, Nanyang Technological University, 637371, Singapore

[4] Aerospace, Mechanical and Mechatronic Engineering, The University of Sydney, 2006, NSW, Australia

[5] Department of Physics & Astronomy, Trent University, 1600 West Bank Dr., Peterborough ON, K9J 0G2, Canada

[6] ARC Centre of Excellence for Transformative Meta-Optical Systems (TMOS), University of Technology Sydney, Ultimo, New South Wales 2007, Australia

**Corresponding**

igor.aharonovich@uts.edu.au; milos.toth@uts.edu.au

**Abstract**

Modifying material properties at the nanoscale is crucially important for devices in nanoelectronics, nanophotonics and quantum information. Optically active defects in wide band gap materials, for instance, are critical constituents for the realisation of quantum technologies. Here, we demonstrate the use of recoil implantation, a method exploiting momentum transfer from accelerated ions, for versatile and mask-free material doping. As a proof of concept, we direct-write arrays of optically active defects into diamond via momentum transfer from a $Xe^+$ focused ion beam (FIB) to thin films of the group IV dopants pre-deposited onto a diamond surface. We further demonstrate the flexibility of the technique, by implanting rare earth ions into the core of a single mode fibre. We conclusively show that the presented technique yields ultra-shallow dopant profiles localised to the top few nanometres of the target surface, and use it to achieve sub-50 nm positional accuracy. The method is applicable to non-planar substrates with complex geometries, and it is suitable for applications such as electronic and magnetic doping of atomically-thin materials and engineering of near-surface states of semiconductor devices.

**Introduction**

Much of solid-state science revolves around modification of materials at the atomic scale, where site-selective control over the incorporation and formation of defects within a host lattice enables control of physical, chemical and optoelectronic properties. Ion implantation has become a key tool in this regard, allowing for the modification of material properties for nanoelectronics, spintronics and quantum photonics.[1-4] Of particular interest is the generation of optically-addressable qubits such as colour centres in diamond, silicon carbide (SiC) or yttrium orthovanadate ($YVO_4$) which are considered prime candidates for scalable quantum technologies.[5-8] Examples include the nitrogen vacancy (NV) centre[9] and more recently the group IV elements[10-16] in diamond that exhibit excellent optical and coherence properties suitable for quantum circuitry.[17-19]

Over the last few decades, significant efforts have been committed to the deterministic creation of colour centres in solid-state host materials by ion implantation.[20-27] While the technique is well established, there are fundamental constraints which limit its application in nanotechnology realizations. For instance, standard ion implanters employ beams that are too wide for direct-write patterning with a precision of tens of nanometres—requiring integration with mask-based lithographic techniques to achieve sub-micron lateral resolution. They are also generally ineffective for the implantation of foreign species into atomically-thin materials such as graphene or transition metal di-chalcogenides,[28] especially when high positional accuracy and lateral resolution is needed. To overcome these intrinsic limitations, focused ion beam (FIB) systems which can operate at low energies and can perform direct writing have been developed. The main drawback of these systems, however, is the narrow range of atomic species available to them: most are limited to elements that can be implanted from a liquid metal ion source.[29]

Here, we demonstrate a feasible method that targets these limits. Stemming from recoil implantation and ion beam mixing,[30-33] the method utilises an inert primary focused ion beam to implant secondary species into a solid host by transfer of momentum. When the method was proposed over half a century ago, ion beam mixing/recoil implantation was mostly studied with a focus on mixing at the interface of adjacent layers, without demonstration of working devices. Nevertheless, the technique combines features which are usually displayed—individually—by different approaches: it works for a large variety of elements, it is mask-free, and it is capable of incorporating dopants with high spatial resolution (~ tens of nanometres) and at shallow depth (~few nanometres). By combining capabilities that typically require a range of instruments and methods, the demonstrated approach to recoil-implantation is complementary to existing ion implanters and FIB instruments, and will broaden their capabilities and reach. It works for a wide variety of chemical species, it is particularly appealing for multi-elemental implantation—both in the context of co-irradiations and serial irradiation—and it can be used on target samples with non-trivial spatial features and geometries. We showcase this versatility by fabricating a range of optically-active, atom-like emitters from solid-state precursors in bulk diamond as well as at the apex of an optical fibre—the latter being a particularly challenging sample for standard mask-based lithographic techniques due its non-planar geometry.

**Results**

**Concept**

To implement the recoil implantation technique, we employ a readily-available, unmodified dual beam system consisting of a coincident scanning electron microscope and a focused ion beam (SEM-FIB) with a xenon (Xe) plasma ion source, as shown in Figure 1a. The doping is performed by site-selective $Xe^+$ beam irradiation of a target that contains the desired dopant species precursor in the form of a thin film. At the atomic scale, momentum transfer from xenon (purple circles in Figure 1a inset) to the precursor atoms (red circles) causes the latter to be implanted into the underlying target. The process can yield ultra-shallow implant profiles, which is desirable for many applications.

As a benchmark for illustrating the versatility of this technique, we choose diamond, with a primary focus on a range of group IV elements in the periodic table: silicon (Si), germanium (Ge), tin (Sn) and lead (Pb). The reasons for the choice of host and dopants are as follows. First, diamond is available as an ultra-pure material with a wide bandgap and exceptional chemical inertness, making it an ideal platform for investigations of extrinsic dopants. Second, colour centres in diamond, particularly the group IV-related negatively charged vacancy complexes, are currently front runner systems for a plethora of quantum photonic and quantum sensing applications.[10] Hence, a robust method to create these centres is desirable. Finally, these defects in diamond are optically active and can be imaged at a single-photon level. This, in turn, allowed us to establish the efficacy of the proposed technique in the extreme case of advanced material systems which require engineering of individual, isolated atomic point defects.

To perform the recoil implantation, the aforementioned group IV elements were deposited onto diamond using a standard sputtering technique (15-nm thickness). Then, a 30 keV, focused $Xe^+$ ion beam was used to pattern the *University of Technology Sydney* logo, and the acronym "UTS" into the thin film sections using an ion beam fluence of $2.5 \times 10^{13}$ $cm^2$. After irradiation, the thin film was stripped using chemical processes and the diamond was annealed to activate the emitters (i.e. vacancy complexes of Pb, Sn, Ge and Si). Further specifications are given in the Methods section and in the Supplementary Note 1. Figure 1b shows a photoluminescence (PL) map of the implanted area. The UTS emblem and the letters U, T, and S are clearly visible. Importantly, the emblem and each of the letters consist of a different dopant—Pb, Sn, Ge, and Si, respectively—implanted using the same inert $Xe^+$ ion beam. We stress that the doping is achieved by momentum transfer from the xenon beam to the aforementioned elements in the thin films deposited atop the diamond. The pattern is formed by electrostatic scanning of the ion beam with no additional lithography steps. The process is performed using a dual FIB-SEM system. All imaging and spatial alignments are therefore performed using the electron beam which is coincident with the ion beam at the diamond surface and serves as a passive imaging tool free from undesired implantation. Figure 1c shows photoluminescence spectra from the respective areas, featuring characteristic emissions of the four colour centres with zero phonon lines (ZPLs) at ≅ 550 nm (PbV), ≅ 620 nm (SnV), ≅ 600 nm (GeV), and ≅ 740 nm (SiV). Remarkably, the colour centres are all optically active and bright, despite the fact that the dopant profiles are ultra-shallow, as is discussed below.

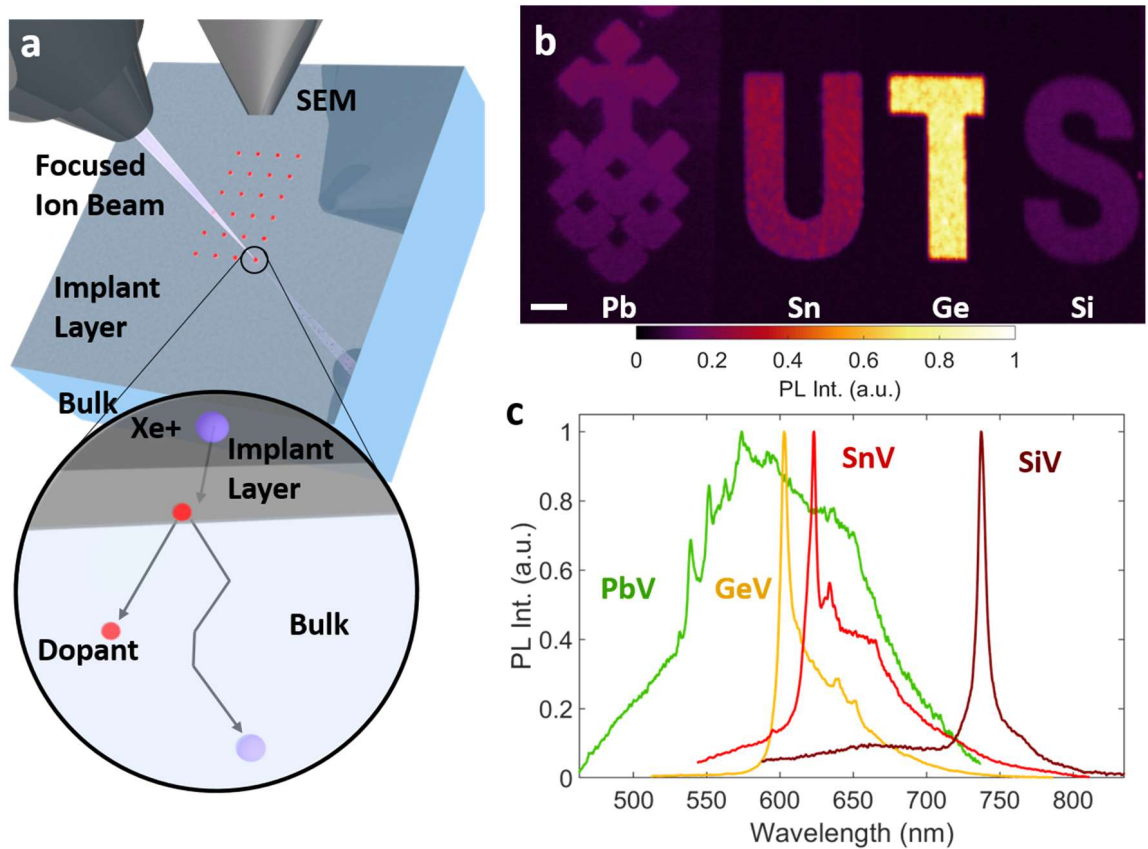

*Figure 1. Creation of colour centres in a diamond target by recoil-implantation. a) Schematic illustration of the process inside a standard electron-ion dual beam microscope. The inset shows an atomic picture of the process: momentum is transferred from inert primary ions to the atoms of a thin film comprised of the dopants, resulting in implantation of the latter into the underlying target. b) Photoluminescence map of a pattern fabricated in a single region of diamond using four dopant species that were implanted by momentum transfer from an electrostatically scanned $Xe^+$ beam. The emblem is patterned in Pb, and the letters U, T and S in Sn, Ge, and Si, respectively. The scale Bar corresponds to 5 µm. c) Photoluminescence spectra recorded from the implants in (b), showing characteristic emissions from defect complexes that contain the Pb (green), Ge (orange), Sn (red) and Si (maroon) dopants.*

**Fluence and spatial control**

We now turn to an in-depth analysis of the technique and the fabricated dopants, starting with the photophysical properties of the emitters. For this purpose, spot arrays (10 × 10) were patterned as a function of Xe$^+$ fluence, which was varied logarithmically for each of the aforementioned dopants. Analysis of the implanted areas was then conducted at room temperature using a confocal setup with an oil-immersion objective. A representative map of a Xe$^+$ fluence-dependent implantation series is shown in Figure 2a. The image refers specifically to the case of silicon-vacancy (SiV) emitters, but identical arrays were also patterned for the other three elements (GeV, PbV and SnV). For all four elements, we observe the emergence of the corresponding colour centre emission beyond a particular ion beam fluence (specified in Figure 2a). We note that the optical signals emerge after an annealing step, which promotes the formation of the fluorescent defect complexes, i.e. impurity atom adjacent to vacancy(ies), displaying photoluminescence emission. The observation of the emission above a particular Xe$^+$ fluence is related to the stochastic nature of the activation process, further discussed in the Supplementary Note 2.

The ion beam fluence needed to form an unambiguous observable pattern, specific to the 15-nm dopant film thickness used in our experiments, is on the order of 2.5×10$^{13}$ cm$^{-2}$, 2.5×10$^{13}$ cm$^{-2}$, 5×10$^{13}$ cm$^{-2}$, and 5×10$^{13}$ cm$^{-2}$ for Si, Ge, Sn and Pb, respectively. This trend with atomic weight is intuitive given that the Xe$^+$ ions mill the 15-nm films via sputtering and simultaneously implant the dopants into the underlying diamond via momentum transfer. Additional details about the creation yield and further characterisation (AFM and PL) are presented and discussed in the Supplementary Note 2 and 3, respectively. We note that co-implanted Xe ions may also form luminescent colour centres in diamond,[34] yet, their spectral features appear at 795 nm and 815 nm and do not overlap spectrally with the studied centres. Moreover, throughout the fluence range used in Figure 2, the Xe-related emission was not observed in PL spectra. Further details are provided in the Supplementary Note 4.

Throughout the implanted arrays (fabricated as a function of Xe$^+$ fluence), prominent ZPLs are observed at 738 nm (SiV), 602 nm (GeV), 620 nm (SnV) and 550 nm (PbV), as shown in Figure 2b, which are characteristic of the respective colour centres. We note that an additional broad peak at 630 nm appearing in some of the spectra is the Raman signal of the immersion oil (Supplementary Note 5). Also, in the case of SnV, an additional peak at 595 nm is present, associated with intermediate defect configurations.[35] To confirm the quantum nature of the emitted light at the lowest implantation fluence, a second order autocorrelation function, ($g^{(2)}(\tau)$), was recorded from each of the emitters. The background-corrected $g^{(2)}(\tau)$ curves are shown as insets in Figure 2b. A clear dip below 0.5 at zero-delay time is observed for GeV, SnV and PbV-related emitters, confirming the single photon nature of the implanted colour centres. We did not observe anti-bunching from the fabricated SiV colour centres, most likely due to their very low quantum efficiency.[36]

The excited state lifetimes of the created defects are in accord with previously reported values—(1.5 ± 0.1) ns, (4.7 ± 0.1) ns, (3.9 ± 0.1) ns, (3.5 ± 0.2) ns, for SiV, GeV, SnV and PbV centres, respectively (Supplementary Note 6).[10] Furthermore, resonant excitation at cryogenic temperatures was performed on some of the emitters. Narrowband linewidths on the order of 5 GHz were observed for the GeV and the SiV colour centres, which is promising for practical applications (Supplementary Note 7). For applications relying on the coherence of the spin state of these defects, further post-processing steps (e.g. high-pressure, high-temperature annealing) may be needed to remove residual crystal damage.

We further analysed the ZPL spectral properties of the ultra-shallow emitters. Interestingly, the distribution of ZPL wavelength increases with the atomic weight of the dopant species, under the exact same annealing conditions. For instance, the SiV ZPL is almost always at the same wavelength ≅ 738 nm, while the PbV and SnV ZPL vary over several nm. Furthermore, additional spectral lines (tentatively attributed to intermediate defect states) appear to a greater extent and with higher frequency for heavier implants, as discussed in the Supplementary Note 8. [14, 35, 37] Whilst some of these characteristics can be attributed to an increase in strain associated with dopant size, these results warrant further investigation, in particular towards optimisation of the annealing process.[35, 38] Nonetheless, the recoil implantation technique is robust and can produce colour centres on demand, as demonstrated here with the group IV family in diamond.

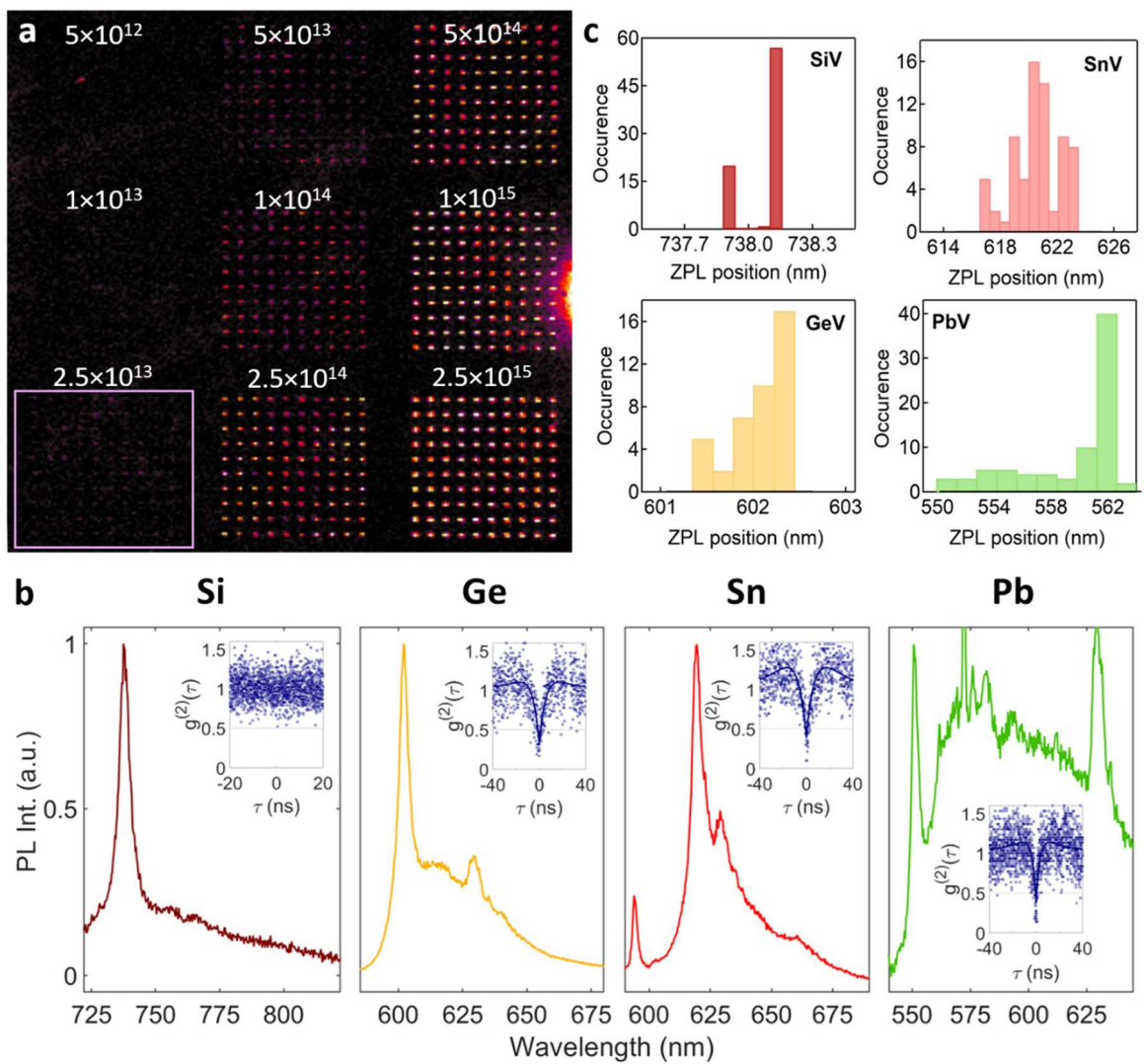

***Figure 2.*** *Spot arrays of dopants implanted into diamond as a function of Xe⁺ fluence. a) Photoluminescence map of a representative array, here shown for Si dopants. The framed area shows the lowest fluence of Xe⁺ that yielded an unambiguous observable array in this case. The number above each array specifies the Xe⁺ fluence in units of cm⁻². b) Photoluminescence spectra of emitters from the lowest observable fluence of SiV (maroon), GeV (orange), SnV (red), and PbV-related emissions (green). An additional peak at 630 nm corresponds to the Raman signal of the utilised immersion oil. Insets show second order correlation measurements (background corrected), illustrating the quantum nature of the colour centres. c) Histograms showing variability in ZPL wavelength in spot arrays of the colour centre ensembles.*

**Spatial control and depth profile**

Next, we analyse the dopant depth profiles, as well as positional accuracy of the recoil implantation. Lateral precision was determined using PL maps of spot-arrays of low fluence ($5 \times 10^{13}$ cm⁻²) GeV ensembles. The ensemble locations were determined by local fits of the PL signal at the nominal implantation sites to 2D gaussian functions and related to a fitted grid, as detailed in the Supplementary Note 9. The distance of the nominal position (i.e. the grid intersections) to the centre of the gaussian fit was calculated, yielding a Rayleigh distribution of the relative ensemble position, as shown in Figure 3a, whereas the extracted relative position did not show any skewness in the x or y direction (inset). A fit to a Rayleigh distribution, yields a mode of (44 ± 4) nm, representative of the implant-placement accuracy of the technique. This value is better than that required for engineering of quantum emitters within photonic devices, e.g. in the high field region of a 2D photonic crystal cavity.[21]

To determine the ultimate lateral resolution of the technique, the above measure of accuracy should be considered in the context of the ion beam diameter, and that of straggle of atoms implanted. An experimental analysis of the ion beam is provided in the Supplementary Note 10. It should be noted that, in the case of the plasma FIB used here, the beam diameter is limited by the virtual source size and the ion beam focusing optics.[39] It can be improved in a number of ways, such as utilisation of a Gas Field Ionisation Source[40], or a pierced cantilever[41] with a hole that is smaller than the beam diameter and thus reduces it in the plane of the target. Beyond the beam diameter, a fundamental limit on resolution is imposed by the radial range and straggle of the implanted atoms. To illustrate the magnitude of this effect, we simulated the spatial distribution of Ge implanted into diamond from a 15-nm film by an ideal 30 keV Xe⁺ beam (i.e., a beam with a diameter of 0 nm). The simulation code and results, detailed in the Supplementary Notes 11 and 12, reveal that the radial range and straggle are approximately 6 and 4 nm, respectively—a minor component of the measured accuracy of (44 ± 4) nm.

The integration of FIB with a coincident high resolution SEM in the same tool enables accurate and localised defect engineering at a particular target site (e.g. the high field region of a photonic resonator). We note that concurrent imaging using the SEM does not lead to any unwanted implantation or alteration of the diamond.

We now discuss the ultra-shallow implant depth profile. Monte Carlo modelling (described in the Supplementary Note 11) indicates that the technique yields dopants within the very top layer of the target surface, with a range and straggle (defined in the Supplementary Note 10) of approximately 0.2 and 0.6 nm, respectively. In more detail, the Ge depth distribution, shown on a log-linear scale in Figure 3b, reveals that over 90% of the implanted species are located within a depth of 1 nm, and the tail extends to approximately 10 nm. This is in good agreement with the measured value of (8 ± 2) nm that is discussed below. For comparison, the depth distribution of the co-implanted Xe is also shown in Figure 3b. The Xe depth profile is much more uniform, and not as concentrated as that of Ge within the top few nanometres of the diamond surface.

Experimentally, we cannot conclusively confirm by direct optical imaging that the emitters are ≤5 nm from the interface. Therefore, to verify the depths of the implanted emitters, we employed Electron Beam Induced Etching (EBIE) in an environmental SEM with $H_2O$ vapour as the etch precursor.[42] EBIE allows us to locally remove the top material of the implanted areas, without damaging the material underneath.[43] During EBIE the electron beam locally induces a chemical reaction with the environmental species ($H_2O$) forming a volatile complex that either desorbs spontaneously or is removed by electron beam induced desorption. The rationale for the deployment of this technique is given by the fact that no further physical alteration of the implanted species is expected, as the etching process is driven chemically at the diamond surface.

Using EBIE, a wedge-shaped feature was etched along a square implanted area ($3.1×10^{12}$ cm$^{-2}$), as is shown schematically in Figure 3c. In the following, we correlated the depth (AFM scan, shown in Figure 3d) to the local fluorescence intensity (PL map, shown in Figure 3e). Using three etched spots (white circles) as fiducial markers to correlate both maps, the average depth and PL profiles of the wedge were extracted (Fig. 3f). In particular, we observe a PL emission maximum from a mean depth of less than $\simeq$1.5 nm in the section—along the x-axis—between the 11 μm and 13 μm mark. We note that the initial increase observable in the PL line profile may be caused by an increase in surface roughness. This hypothesis is supported by the observation that the increase in PL correlates with spot-like features clearly discernible in the PL map in Figure 3e.

Consequently, as the mean depth drops below 1.5 nm, the PL intensity decreases sharply to less than 50%, in the section along the x-axis, $x \cong$ 13–14 μm and down to 20% at about (5 ± 2) nm at $x \cong$ 15 μm. However, a residual PL signal is still present at a depth of (8 ± 2) nm, which is in good agreement with the tail of the simulated depth profile (Fig 3b). Specifically, at a depth of $\cong$ 8 nm, the number of Ge atoms drops below 1, which in combination with a conversion efficiency on the order of $\cong$ 1%[22] make it unlikely that colour centres are created beyond this depth. We note that the simulations do not account for channelling, which is likely insignificant, as indicated by an analysis of the simulated Ge atom momentum distribution at the film-diamond interface. Specifically, the momentum analysis shown in the Supplementary Note 12 reveals that the strong forward-directionality of the Xe$^+$ beam is not retained by the Ge atoms.

Returning to Figure 3f, the luminescence signal is restored at $x \cong 22$ μm, corresponding to the end of the etched wedge. As one progresses further along the x-axis, the luminescence drops again to zero at the edge of the implanted area at $x \cong 25$ μm.

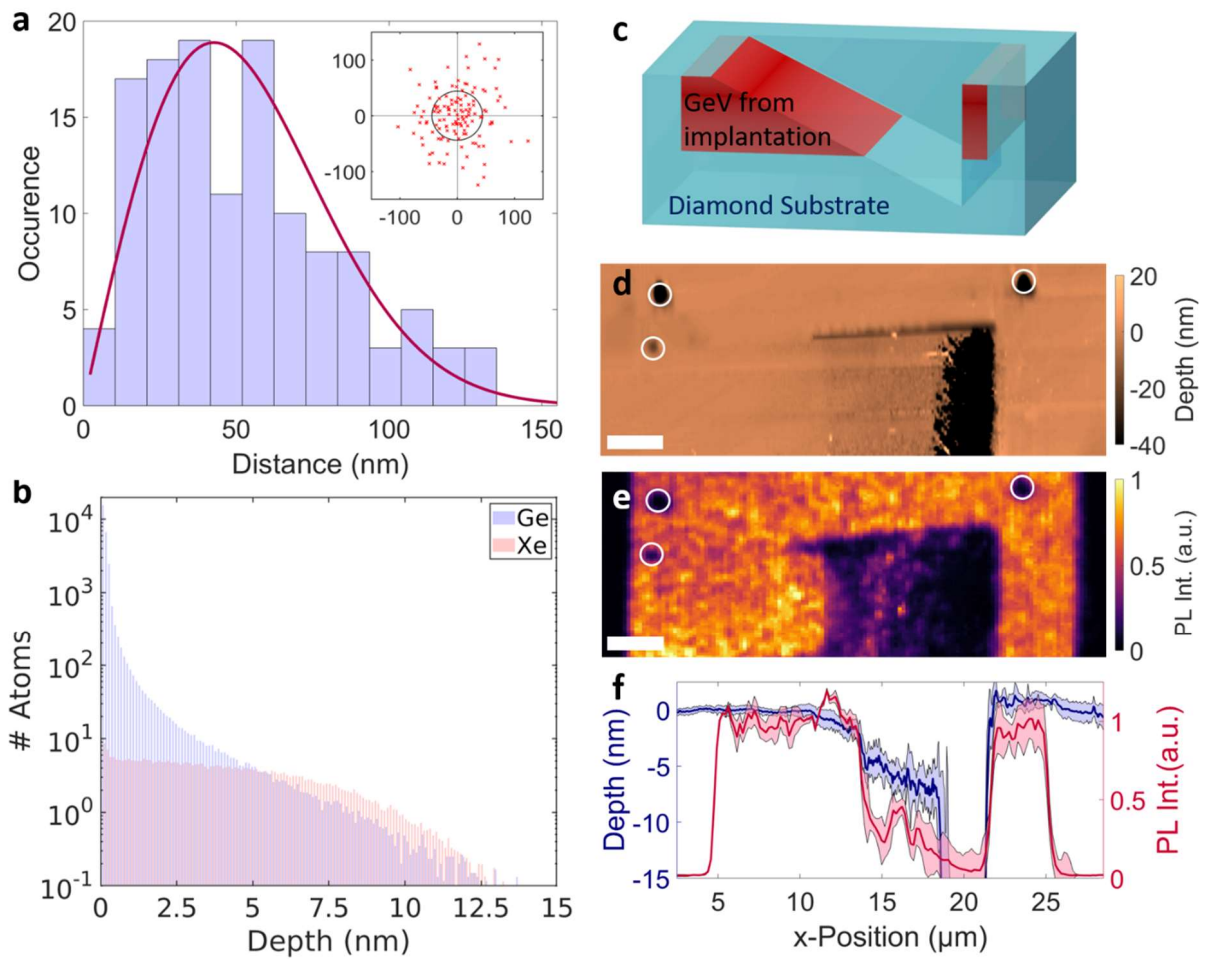

*Figure 3. Spatial distribution of the dopants. a) Distribution of the ensemble position (GeV) relative to the nominal implantation site, fitted by a Rayleigh function with a mode of (44 ± 4) nm. The inset shows directly the relative ensemble position. b) Monte Carlo simulations of the depth distribution of Ge in diamond from a Xe$^+$ recoil-implantation process and of the co-implanted Xe. c) Schematic of the wedge edge to access the implantation profile of an area with the implanted species. d) AFM map of the wedge prepared by EBIE. e) PL map of the same Ge-implanted area. The positions of both maps were correlated by the 3 marked spots. Scale bars in (d) and (e) correspond to 2.5 μm. f) Average profile of depth (blue) and PL (red) along the wedge feature. The shaded areas represent the standard deviation.*

**Implantation into restrictive geometries**

Next, we demonstrate another key feature of the recoil implantation technique by implanting a specific site of a target that has a restrictive geometry. Most common implantation methods do not use focused ion beams, but instead achieve site-specificity using mask-based lithographic techniques that are often inapplicable to relatively small, high aspect ratio, non-planar samples. To showcase how this constraint is overcome in an extreme example, we implanted a rare earth element, Europium, into the end face of a commercial single-mode optical fibre (Figure 4a). We choose Europium to emphasise three key features. 1) Europium is heavier than xenon, which together with the Pb data in Figure 1 and 2 shows that the technique is applicable to most elements in the periodic table. 2) The implantation of Eu at low energies is challenging by conventional methods since it is difficult to form the stable negative Eu ions needed by some ion implanters.[44] 3) The integration of rare earth elements with optical fibres is generally sought after for integrated photonics.

Analogous to the above experiments, a thin film of Europium was deposited on the fibre end. The fibre was then mounted in the FIB-SEM dual beam microscope and the fibre core was identified using the SEM, as shown in Figure 4b. The inset of figure 4b is an optical image of the fibre, with the core lit up in blue by back-flowing light from a flashlight through the opposite end of the fibre. The mark "x" is a fiducial marker used to correlate the optical and SEM images. To implant the Eu, the core of the fibre was irradiated using a $Xe^+$ fluence of $4.1 \times 10^{13}$ $cm^{-2}$. This value was determined beforehand by implantation into $SiO_2$ (see Supplementary Note 13). Subsequently, the thin film was removed and the implanted area was investigated by PL analysis. As shown in Figure 4c, we observed the emission related to the $^5D_0 \rightarrow {}^7F_2$ transition of the $Eu^{3+}$ ion centred at a wavelength of $\cong$ 620 nm. The emission is confined to the core of the fibre, as shown by the confocal PL map obtained by exciting the $Eu^{3+}$ from the back of the fibre.

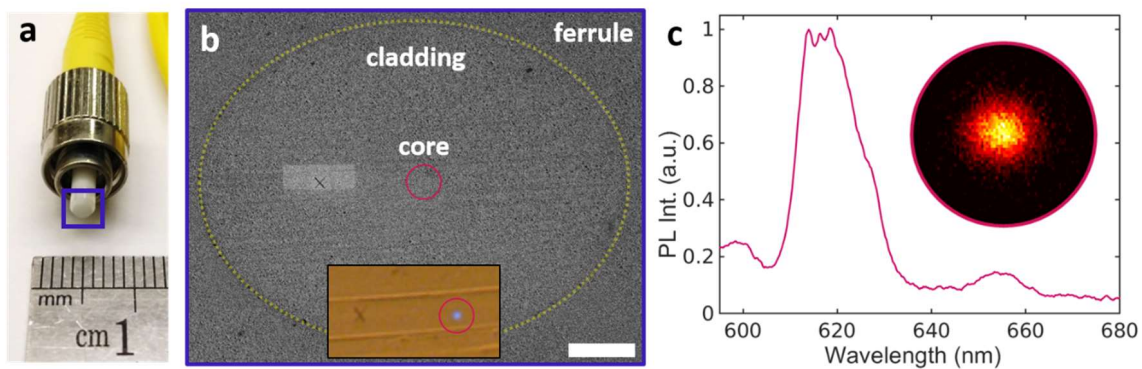

*Figure 4. Implantation of rare earth ions into a commercial optical fibre. a) Picture of the fibre used in this experiment. The end face is highlighted. b) SEM image of the fibre end face during the implantation process. The yellow outline marks the boundary of the cladding to the ferrule, which was used to identify the core. The inset shows the core (back illuminated) relative to a correlation marker. The scale bar corresponds to 20 µm. c) Emission of the implanted $Eu^{3+}$, excited through the other fibre end (PL map shown as inset).*

**Discussion**

To conclude, we demonstrated versatile creation of shallow implants in a solid-sate host using recoil implantation. This is achieved by utilizing momentum transfer from a $Xe^+$ FIB to atoms of a thin target film deposited on top of a target of choice. In our work this is demonstrated by implantation from solid thin films deposited on the sample surface (diamond in our case). A compelling aspect of the technique is the ability to achieve ultra-shallow implants. We used the method to create optically-active, atom-like defects in diamond and demonstrated their formation within the top 6 nm from the surface. To demonstrate the applicability of the technique to non conventional samples, we engineered rare earth elements directly in the core of a single mode optical fibre. Beyond the field of quantum technologies, the recoil implantation technique is appealing for shallow doping of single electron transistors,[45] deterministic direct writing of near-surface dopants[46, 47] or controlled introduction of magnetic elements for magnetism at the nanoscale.[48] Finally, one of the biggest advantages of this technique is the ability to engineer selected defects in atomically-thin materials, which by definition requires ultra-shallow implants.

**Methods**

**Sample Preparation**

A CVD-grown electronic grade diamond (< 1ppb Nitrogen) was purchased from Element Six. Before all subsequent experiments were carried out, the sample was cleaned in hot (150 °C) Piranha Acid ($H_2SO_4$:$H_2O_2$ (30%) 2:1) for at least 2 hours. All thin film depositions were carried out in a lab-built magnetron sputter deposition chamber, pumped to a pressure of $1\times10^{-5}$ Torr or lower before deposition. For deposition, argon was introduced with a pressure of $1.5\times10^{-3}$ Torr and a plasma was ignited, for which the power was adjusted to yield a deposition rate of 0.5 Å $s^{-1}$. For the implantation experiments discussed in Figure 2 and Figure 3 the diamond was coated at 4 different corners with the respective thin films, whilst being masked during subsequent deposition steps to avoid cross deposition of different materials. For the patterning of the UTS logo, thin films were deposited on rectangular sections, lithographically defined using EBL as described in Supplementary Note 1. For implantation into an optical fibre or $SiO_2$, a 15-nm Eu film was deposited on top in a thermal evaporator using Eu pellets. Sputter targets and evaporation material were purchased from Changsha Xinkang Advanced Materials Co., Ltd with a purity of 99.99 % or higher.

**Focused Ion Beam Irradiation**

All ion beam irradiations were carried out in a Thermo Fisher Scientific Helios G4 PFIB with exchangeable plasma ion source. For irradiation, a xenon beam was used with an acceleration voltage of 30 kV and current of 10 pA. For irradiation of spot arrays and the UTS logo, custom patterning files (stream files) were created, defining the time per spot for individual spots (pixels). Box irradiations as discussed in the Supplementary Note 3 were performed, defining squares with 4 µm length and increasing the number of passes during the irradiation, using the built-in pattern generator.

**Post Irradiation treatment**

For experiments on diamond, after irradiation with the $Xe^+$ FIB, the coated thin films were chemically stripped. We used subsequently 2 cycles of various solutions with intermittent water rinsing in between. First, the sample was held in KOH (30 min., 50 °C, 30 wt. %), then HCl (30 min., 75 °C, 37 wt. %), followed by Piranha Acid (30 min, 150 °C $H_2SO_4$:$H_2O_2$ (30%) 2:1) to efficiently eliminate any residue of thin film coating on the substrate surface. After the chemical treatments the sample was annealed in a tube furnace (Lindberg Blue Mini-mite) under high vacuum with pressure lower than $2\times10^{-6}$ Torr for the entire annealing cycle. The temperature was ramped up to 950 °C and held for 2 hours. Then the furnace was cooled down to room temperature before breaking vacuum. After annealing the sample was again cleaned in hot (150 °C) Piranha Acid ($H_2SO_4$:$H_2O_2$ (30%) 2:1) for at least 2 hours and stored in a desiccator in between experiments. For experiments on $SiO_2$ and the optical fibre, the Eu thin film was removed after implantation in warm HCl 50 °C and subsequently water. No annealing step was performed to activate the Eu3+ related emission.

**PL measurements**

All PL measurements were performed on lab-built confocal setups. The UTS logo was characterised using a 405-nm, 1mW cw laser as the excitation source, which was directed through an arrangement of a dichroic mirror, scanning mirror and lens relay system and focused on the sample surface by a 0.9 NA air objective. Photoluminescence from the sample was collected from the same objective and directed into a multimode fibre guiding the signal either to an APD (excelitas) for mapping or to a spectrometer (Princeton Instruments). Lifetime measurements were conducted on the same configuration using either a 512 nm (for Si, Ge, Sn) or 405 nm (for Pb) pulsed laser diode as the excitation source. The fluorescence was correlated to the laser pulse using a correlator (Pico Harp 300, Picoquant). For the characterisation of the fibre core, the fibre was mounted in front of the air objective and a 2 mW, 532-nm cw laser was connected to the unimplanted end of the fibre to excite implanted $Eu^{3+}$ in the core on the other end. At the same time, the collection was scanned in the same configuration. The characterisation of spot arrays and second order correlation measurements were done on a different setup. A 532-nm laser (cw, 2 mW) was used as the excitation source, and focused onto the sample by using an oil immersion objective (1.3 NA). Photoluminescence was collected through the same objective and coupled into a single mode fibre. The signal was either guided to a Spectrometer (Andor), or a 50/50 fibre splitter with the respective outputs connected to an APD. Their signals were then correlated by a time correlator (Swabian Instruments).

**Electron Beam Induced Etching**

Electron beam induced etching of the diamond was done in a Zeiss EVO, under 15 keV, 2 nA, 100 mTorr $H_2O$ vapour. A rectangular area (16 µm length, 10 µm width) was irradiated (5 µs dwell time, 20 nm point pitch) in a raster scan. The length of the rectangle was reduced in subsequent steps by 2 μm, whilst the time/ area (≈1 s·μm$^{-1}$) for each rectangle was held constant, thus achieving a linear fluence gradient resulting in the shown etch profile. After etching the diamond was again cleaned in hot (150 °C) Piranha Acid ($H_2SO_4$:$H_2O_2$ (30%) 2:1).

**AFM measurement**

Atomic force microscopy (AFM) measurements were done on a Park XE-7 AFM. Post-processing of AFM data was done in XEI and Gwyddion[49] to remove artefacts of the scan and to extract height profiles.

**References**

1. Hensen, B.; Wei Huang, W.; Yang, C.-H.; Wai Chan, K.; Yoneda, J.; Tanttu, T.; Hudson, F. E.; Laucht, A.; Itoh, K. M.; Ladd, T. D.; Morello, A.; Dzurak, A. S., A Silicon Quantum-Dot-Coupled Nuclear Spin Qubit. *Nat. Nanotechnol.* **2020**, *15*, 13-17.
2. van Donkelaar, J.; Yang, C.; Alves, A. D. C.; McCallum, J. C.; Hougaard, C.; Johnson, B. C.; Hudson, F. E.; Dzurak, A. S.; Morello, A.; Spemann, D.; Jamieson, D. N., Single Atom Devices by Ion Implantation. *Journal of Physics: Condensed Matter* **2015**, *27*, 154204.
3. Tuček, J.; Błoński, P.; Ugolotti, J.; Swain, A. K.; Enoki, T.; Zbořil, R., Emerging Chemical Strategies for Imprinting Magnetism in Graphene and Related 2d Materials for Spintronic and Biomedical Applications. *Chemical Society Reviews* **2018**, *47*, 3899-3990.
4. Awschalom, D. D.; Hanson, R.; Wrachtrup, J.; Zhou, B. B., Quantum Technologies with Optically Interfaced Solid-State Spins. *Nat. Photonics* **2018**, *12*, 516-527.
5. Atatüre, M.; Englund, D.; Vamivakas, N.; Lee, S.-Y.; Wrachtrup, J., Material Platforms for Spin-Based Photonic Quantum Technologies. *Nat. Rev. Mater.* **2018**, *3*, 38-51.
6. Mouradian, S. L.; Schröder, T.; Poitras, C. B.; Li, L.; Goldstein, J.; Chen, E. H.; Walsh, M.; Cardenas, J.; Markham, M. L.; Twitchen, D. J.; Lipson, M.; Englund, D., Scalable Integration of Long-Lived Quantum Memories into a Photonic Circuit. *Phys. Rev. X* **2015**, *5*, 031009.
7. Lukin, D. M.; Dory, C.; Guidry, M. A.; Yang, K. Y.; Mishra, S. D.; Trivedi, R.; Radulaski, M.; Sun, S.; Vercruysse, D.; Ahn, G. H.; Vučković, J., 4h-Silicon-Carbide-on-Insulator for Integrated Quantum and Nonlinear Photonics. *Nat. Photonics* **2019**.
8. Zhong, T.; Kindem, J. M.; Bartholomew, J. G.; Rochman, J.; Craiciu, I.; Miyazono, E.; Bettinelli, M.; Cavalli, E.; Verma, V.; Nam, S. W.; Marsili, F.; Shaw, M. D.; Beyer, A. D.; Faraon, A., Nanophotonic Rare-Earth Quantum Memory with Optically Controlled Retrieval. *Science* **2017**, *357*, 1392.
9. Doherty, M. W.; Manson, N. B.; Delaney, P.; Jelezko, F.; Wrachtrup, J.; Hollenberg, L. C. L., The Nitrogen-Vacancy Colour Centre in Diamond. *Physics Reports* **2013**, *528*, 1-45.
10. Bradac, C.; Gao, W.; Forneris, J.; Trusheim, M. E.; Aharonovich, I., Quantum Nanophotonics with Group Iv Defects in Diamond. *Nat. Commun.* **2019**, *10*, 5625.
11. Neu, E.; Steinmetz, D.; Riedrich-Moeller, J.; Gsell, S.; Fischer, M.; Schreck, M.; Becher, C., Single Photon Emission from Silicon-Vacancy Centres in Cvd-Nano-Diamonds on Iridium *New Journal of Physics* **2011**, *13*, 025012.
12. Iwasaki, T.; Ishibashi, F.; Miyamoto, Y.; Doi, Y.; Kobayashi, S.; Miyazaki, T.; Tahara, K.; Jahnke, K. D.; Rogers, L. J.; Naydenov, B.; Jelezko, F.; Yamasaki, S.; Nagamachi, S.; Inubushi, T.; Mizuochi, N.; Hatano, M., Germanium-Vacancy Single Color Centers in Diamond. *Sci. Rep.* **2015**, *5*, 12882.
13. Johannes, G.; Dennis, H.; Gergö, T.; Philipp, F.; Morgane, G.; Takayuki, I.; Takashi, T.; Michael, K.; Meijer, J.; Mutsuko, H.; Adam, G.; Christoph, B., Spectroscopic Investigations of Negatively Charged Tin-Vacancy Centres in Diamond. *New Journal of Physics* **2019**.

14. Trusheim, M. E.; Wan, N. H.; Chen, K. C.; Ciccarino, C. J.; Flick, J.; Sundararaman, R.; Malladi, G.; Bersin, E.; Walsh, M.; Lienhard, B.; Bakhru, H.; Narang, P.; Englund, D., Lead-Related Quantum Emitters in Diamond. *Physical Review B* **2019**, *99*, 075430.
15. Ditalia Tchernij, S.; Lühmann, T.; Herzig, T.; Küpper, J.; Damin, A.; Santonocito, S.; Signorile, M.; Traina, P.; Moreva, E.; Celegato, F.; Pezzagna, S.; Degiovanni, I. P.; Olivero, P.; Jakšić, M.; Meijer, J.; Genovese, P. M.; Forneris, J., Single-Photon Emitters in Lead-Implanted Single-Crystal Diamond. *ACS Photonics* **2018**, *5*, 4864-4871.
16. Tchernij, S. D.; Herzig, T.; Forneris, J.; Küpper, J.; Pezzagna, S.; Traina, P.; Moreva, E.; Degiovanni, I. P.; Brida, G.; Skukan, N.; Genovese, M.; Jakšić, M.; Meijer, J.; Olivero, P., Single-Photon-Emitting Optical Centers in Diamond Fabricated Upon Sn Implantation. *ACS Photonics* **2017**, *4*, 2580-2586.
17. Trusheim, M. E.; Pingault, B.; Wan, N. H.; Gündoğan, M.; De Santis, L.; Debroux, R.; Gangloff, D.; Purser, C.; Chen, K. C.; Walsh, M.; Rose, J. J.; Becker, J. N.; Lienhard, B.; Bersin, E.; Paradeisanos, I.; Wang, G.; Lyzwa, D.; Montblanch, A. R. P.; Malladi, G.; Bakhru, H.*, et al.*, Transform-Limited Photons from a Coherent Tin-Vacancy Spin in Diamond. *Phys. Rev. Lett.* **2020**, *124*, 023602.
18. Sohn, Y.-I.; Meesala, S.; Pingault, B.; Atikian, H. A.; Holzgrafe, J.; Gündoğan, M.; Stavrakas, C.; Stanley, M. J.; Sipahigil, A.; Choi, J.; Zhang, M.; Pacheco, J. L.; Abraham, J.; Bielejec, E.; Lukin, M. D.; Atatüre, M.; Lončar, M., Controlling the Coherence of a Diamond Spin Qubit through Its Strain Environment. *Nat. Commun.* **2018**, *9*, 2012.
19. Sipahigil, A.; Evans, R. E.; Sukachev, D. D.; Burek, M. J.; Borregaard, J.; Bhaskar, M. K.; Nguyen, C. T.; Pacheco, J. L.; Atikian, H. A.; Meuwly, C.; Camacho, R. M.; Jelezko, F.; Bielejec, E.; Park, H.; Lončar, M.; Lukin, M. D., An Integrated Diamond Nanophotonics Platform for Quantum Optical Networks. *Science* **2016**, *354*, 847.
20. Scarabelli, D.; Trusheim, M.; Gaathon, O.; Englund, D.; Wind, S. J., Nanoscale Engineering of Closely-Spaced Electronic Spins in Diamond. *Nano Lett.* **2016**, *16*, 4982-4990.
21. Schröder, T.; Trusheim, M. E.; Walsh, M.; Li, L.; Zheng, J.; Schukraft, M.; Sipahigil, A.; Evans, R. E.; Sukachev, D. D.; Nguyen, C. T.; Pacheco, J. L.; Camacho, R. M.; Bielejec, E. S.; Lukin, M. D.; Englund, D., Scalable Focused Ion Beam Creation of Nearly Lifetime-Limited Single Quantum Emitters in Diamond Nanostructures. *Nat. Commun.* **2017**, *8*, 15376.
22. Zhou, Y.; Mu, Z.; Adamo, G.; Bauerdick, S.; Rudzinski, A.; Aharonovich, I.; Gao, W.-b., Direct Writing of Single Germanium Vacancy Center Arrays in Diamond. *New Journal of Physics* **2018**, *20*, 125004.
23. Lühmann, T.; John, R.; Wunderlich, R.; Meijer, J.; Pezzagna, S., Coulomb-Driven Single Defect Engineering for Scalable Qubits and Spin Sensors in Diamond. *Nat. Commun.* **2019**, *10*, 4956.
24. Haruyama, M.; Onoda, S.; Higuchi, T.; Kada, W.; Chiba, A.; Hirano, Y.; Teraji, T.; Igarashi, R.; Kawai, S.; Kawarada, H.; Ishii, Y.; Fukuda, R.; Tanii, T.; Isoya, J.; Ohshima, T.; Hanaizumi, O., Triple Nitrogen-Vacancy Centre Fabrication by C5n4hn Ion Implantation. *Nat. Commun.* **2019**, *10*, 2664.
25. Wang, J.; Zhou, Y.; Zhang, X.; Liu, F.; Li, Y.; Li, K.; Liu, Z.; Wang, G.; Gao, W., Efficient Generation of an Array of Single Silicon-Vacancy Defects in Silicon Carbide. *Physical Review Applied* **2017**, *7*, 064021.
26. Groot-Berning, K.; Kornher, T.; Jacob, G.; Stopp, F.; Dawkins, S. T.; Kolesov, R.; Wrachtrup, J.; Singer, K.; Schmidt-Kaler, F., Deterministic Single-Ion Implantation of Rare-Earth Ions for Nanometer-Resolution Color-Center Generation. *Phys. Rev. Lett.* **2019**, *123*, 106802.
27. Siyushev, P.; Xia, K.; Reuter, R.; Jamali, M.; Zhao, N.; Yang, N.; Duan, C.; Kukharchyk, N.; Wieck, A. D.; Kolesov, R.; Wrachtrup, J., Coherent Properties of Single Rare-Earth Spin Qubits. *Nat. Commun.* **2014**, *5*, 3895.
28. Bangert, U.; Pierce, W.; Kepaptsoglou, D. M.; Ramasse, Q.; Zan, R.; Gass, M. H.; Van den Berg, J. A.; Boothroyd, C. B.; Amani, J.; Hofsäss, H., Ion Implantation of Graphene—toward Ic Compatible Technologies. *Nano Lett.* **2013**, *13*, 4902-4907.

29. Bischoff, L.; Mazarov, P.; Bruchhaus, L.; Gierak, J., Liquid Metal Alloy Ion Sources—an Alternative for Focussed Ion Beam Technology. *Applied Physics Reviews* **2016,** *3*, 021101.
30. Nelson, R. S., The Theory of Recoil Implantation. *Radiation Effects* **1969,** *2*, 47-50.
31. Paine, B. M.; Averback, R. S., Ion Beam Mixing: Basic Experiments. *Nuclear Instruments and Methods in Physics Research Section B: Beam Interactions with Materials and Atoms* **1985,** *7-8*, 666-675.
32. Möller, W.; Eckstein, W., Ion Mixing and Recoil Implantation Simulations by Means of Tridyn. *Nuclear Instruments and Methods in Physics Research Section B: Beam Interactions with Materials and Atoms* **1985,** *7-8*, 645-649.
33. Sigmund, P., Recoil Implantation and Ion-Beam-Induced Composition Changes in Alloys and Compounds. *Journal of Applied Physics* **1979,** *50*, 7261-7263.
34. Sandstrom, R.; Ke, L.; Martin, A.; Wang, Z.; Kianinia, M.; Green, B.; Gao, W.-b.; Aharonovich, I., Optical Properties of Implanted Xe Color Centers in Diamond. *Optics Communications* **2018,** *411*, 182-186.
35. Iwasaki, T.; Miyamoto, Y.; Taniguchi, T.; Siyushev, P.; Metsch, M. H.; Jelezko, F.; Hatano, M., Tin-Vacancy Quantum Emitters in Diamond. *Phys. Rev. Lett.* **2017,** *119*, 253601.
36. Tamura, S.; Koike, G.; Komatsubara, A.; Teraji, T.; Onoda, S.; McGuinness, L. P.; Rogers, L.; Naydenov, B.; Wu, E.; Yan, L.; Jelezko, F.; Ohshima, T.; Isoya, J.; Shinada, T.; Tanii, T., Array of Bright Silicon-Vacancy Centers in Diamond Fabricated by Low-Energy Focused Ion Beam Implantation. *Applied Physics Express* **2014,** *7*, 115201.
37. Lindner, S.; Bommer, A.; Muzha, A.; Krueger, A.; Gines, L.; Mandal, S.; Williams, O.; Londero, E.; Gali, A.; Becher, C., Strongly Inhomogeneous Distribution of Spectral Properties of Silicon-Vacancy Color Centers in Nanodiamonds. *New Journal of Physics* **2018,** *20*, 115002.
38. Rugar, A. E.; Lu, H.; Dory, C.; Sun, S.; McQuade, P. J.; Shen, Z.-X.; Melosh, N. A.; Vučković, J., Generation of Tin-Vacancy Centers in Diamond Via Shallow Ion Implantation and Subsequent Diamond Overgrowth. *Nano Lett.* **2020,** *20*, 1614-1619.
39. Burnett, T. L.; Kelley, R.; Winiarski, B.; Contreras, L.; Daly, M.; Gholinia, A.; Burke, M. G.; Withers, P. J., Large Volume Serial Section Tomography by Xe Plasma Fib Dual Beam Microscopy. *Ultramicroscopy* **2016,** *161*, 119-129.
40. Bassim, N.; Scott, K.; Giannuzzi, L. A., Recent Advances in Focused Ion Beam Technology and Applications. *MRS Bulletin* **2014,** *39*, 317-325.
41. Pezzagna, S.; Wildanger, D.; Mazarov, P.; Wieck, A. D.; Sarov, Y.; Rangeow, I.; Naydenov, B.; Jelezko, F.; Hell, S. W.; Meijer, J., Nanoscale Engineering and Optical Addressing of Single Spins in Diamond. *Small* **2010,** *6*, 2117-2121.
42. Bishop, J.; Fronzi, M.; Elbadawi, C.; Nikam, V.; Pritchard, J.; Fröch, J. E.; Duong, N. M. H.; Ford, M. J.; Aharonovich, I.; Lobo, C. J.; Toth, M., Deterministic Nanopatterning of Diamond Using Electron Beams. *ACS Nano* **2018,** *12*, 2873-2882.
43. Martin, A. A.; Toth, M.; Aharonovich, I., Subtractive 3d Printing of Optically Active Diamond Structures. *Sci. Rep.* **2014,** *4*, 5022.
44. Middleton, R. J. U. o. P., unpublished, A Negative Ion Cookbook. **1989**.
45. Koch, M.; Keizer, J. G.; Pakkiam, P.; Keith, D.; House, M. G.; Peretz, E.; Simmons, M. Y., Spin Read-out in Atomic Qubits in an All-Epitaxial Three-Dimensional Transistor. *Nat. Nanotechnol.* **2019,** *14*, 137-140.
46. Bali, R.; Wintz, S.; Meutzner, F.; Hübner, R.; Boucher, R.; Ünal, A. A.; Valencia, S.; Neudert, A.; Potzger, K.; Bauch, J.; Kronast, F.; Facsko, S.; Lindner, J.; Fassbender, J., Printing Nearly-Discrete Magnetic Patterns Using Chemical Disorder Induced Ferromagnetism. *Nano Lett.* **2014,** *14*, 435-441.
47. Hughes, M. A.; Fedorenko, Y.; Gholipour, B.; Yao, J.; Lee, T.-H.; Gwilliam, R. M.; Homewood, K. P.; Hinder, S.; Hewak, D. W.; Elliott, S. R.; Curry, R. J., N-Type Chalcogenides by Ion Implantation. *Nat. Commun.* **2014,** *5*, 5346.

48. Fei, Z.; Huang, B.; Malinowski, P.; Wang, W.; Song, T.; Sanchez, J.; Yao, W.; Xiao, D.; Zhu, X.; May, A. F.; Wu, W.; Cobden, D. H.; Chu, J.-H.; Xu, X., Two-Dimensional Itinerant Ferromagnetism in Atomically Thin Fe3gete2. *Nature Materials* **2018,** *17*, 778-782.
49. Nečas, D.; Klapetek, P., Gwyddion: An Open-Source Software for Spm Data Analysis. *Central European Journal of Physics* **2012,** *10*, 181-188.

# Author Correspondence to the Editor

08/05/2020

Re: "Knock-on doping: A universal method to direct-write dopants in a solid state host"

Dear Dr ___,

Thank you for sending us the referee reports and for giving us the opportunity to revise the work. We are very pleased with the overall positive and encouraging comments from the reviewers and the overall support for publication in n_xxxxxx.

We took all their comments on board, revised the manuscript and added three significant changes:

1. We performed an additional unique knock-on implant into an end facet of the fibre. Such an experiment would be tremendously complicated using commercial implants (impossible really), but is easy using the proposed technique. The data is in figure 4.
2. We added a more rigorous calculations on the ion struggle and beam diameter to convey the ultra shallow claims. The data is in figure S14.
3. We performed additional unique implant from a gaseous phase, implanting nitrogen (i.e. knock on implant of nitrogen). This was noted as "impossible" by referee #3. The data is in figure S3.

We would also like to reiterate a key message about our work. One of the reviewers commented about implanting challenges of lithium while others stated that focused low energy implants are not a problem. It is important to put these things in perspective. For example, this company http://cuttingedgeions.com/Species_List.html which is the most used by researchers in the diamond community, can easily implant lithium, but they cant, for instance, implant many other elements (see the list in that link). On the other hand, while we acknowledge that selected laboratories can do focused nitrogen implantations, implanting group III elements at low energies is extremely challenging, as they don't form good negative ions. This is the case for instance at the ANU facilities https://physics.anu.edu.au/eme/capabilities/implantation.php to which referee 3 was referring. Our method circumvent these issues to provide a truly universal approach for solid state doping, as we now clarified throughout the paper.

We trust the revised version addresses all the concerns raised by the referee and look forward for its acceptance to n_xxxxxx.

Prof Igor Aharonovich

Reviewer #1 (Remarks to the Author):

This interesting article reports about an innovative and undoubtedly appealing experimental approach for the single-defect engineering in solid state, with specific implications in quantum technologies. Despite the potential drawback represented by the co-implantation of primary ions (see below for more details), the reported results are very relevant in consideration of the current state of the art and lucidly presented. Therefore the article is considered as publishable, provided that the following points are suitably clarified by the authors.

*Authors: We thank the Reviewer for acknowledging our work and the general recommendation for publication in N_xxxxxx. The points raised by the Reviewer were indeed helpful and helped improve the quality of the manuscript.*

One of the key features that could potentially limit the applicability of this (otherwise very appealing) technique is that the dynamics of the energy transfer in the knock-on collisions between the primary ions and the target atoms in the deposited films is such that it is impossible to have the secondary atoms implanted in the substrate without the concurrent implantation of the primary ions. To some extent, this effect can be mitigated by employing (such as in the present work) ion beams from inert atomic species. Nonetheless, the authors should not disregard that i) according to several accounts (which should be suitably cited), even inert atomic species seem to yield to the formation of stable optically-active defects in diamond, and ii) more generally, "collateral" structural damage can be detrimental to the spin properties of the defects. Overall, this drawback does not prevent the publishability of this interesting report, but it should be more explicitly assessed in the conclusions.

*Authors: We added a discussion and references explaining that inert ion bombardment can generate colour centers. We also added experimental data to the Supplementary Information 5 to demonstrate that $Xe^+$ ion beam irradiation can indeed generate colour centers in diamond under some conditions.*

*"...We note that co-implanted Xe ions may also form luminescent colour centres in diamond, yet, their spectral features appear at 795 nm and 815 nm and do not overlap spectrally with the studied centres. Moreover, throughout the fluence range used in Figure 2, the Xe-related emission was not observable in PL spectra. Further details are provided in the Supporting Information 6. ..."*

*We also added a discussion addressing the second point raised by the reviewer - that collateral structural damage can affect the spin properties of defects. This can be mitigated by appropriate post-processing of the target.*

*"...For applications relying on the coherence of the spin state of these defects, further post-processing steps (e.g. high-pressure, high-temperature annealing) may be needed to remove residual crystal damage. ..."*

"However, the technique is inherently problematic. [...]": the following points are substantially acceptable, but rather than on cost / technical complexity / availability (a state-of-the-art Xe-FIB is arguably no much cheaper, maintenance-friendly or readily available than a conventional ion implanter) and ion penetration depth (one can always consider masking techniques or counter-fields to slow down the ions) I would suggest focusing more the argument (and thus the relevance of this technique) on the third point, i.e. the readiness in access to multi-elemental implantations, both in the context of co-irradiations and subsequent serial irradiations. I would suggest elaborating more on this point, also because the statement "due to their need for long pumping times and the large spot size of the beam" could be more clearly articulated.

Authors: We revised the introduction and removed the points raised by the Reviewer or clarified them. In detail, we now emphasize the readiness of the technique in the context of co-irradiations and serial irradiations. Moreover, we added new experimental data to further highlight the readiness, universality and versatility of the technique. Specifically, we showed that dopants from vapor-phase precursors can be delivered to the target and implanted by the $Xe^+$ ion beam. We demonstrated this using a capillary-style gas injector, and used this approach to fabricate nitrogen-vacancy (NV) colour centres by $Xe^+$ ion beam irradiation of diamond in the presence of gas-phase $N_2$. These new results show that our technique is applicable not only to solid-state precursors supplied in the form of a thin film, but also to gases, as well as liquids and solids whose vapors can be delivered to the target using gas injection methods that are available on most commercially-available FIB instruments (as is now explained in the revised manuscript and supporting references). The technique is thus readily applicable to essentially all isotopically-stable elements.

We also note that the vapour-phase approach yields an additional benefit for direct-write, localized doping, because film deposition/removal steps are not necessary.

We de-emphasized and clarified our claims about cost-benefit. Our method features the combination of high lateral resolution (without the need for mask-based lithography), shallow implantation depth and extremely high dopant species flexibility. The cost of an equivalent setup based on conventional implanters and mask-based lithographic tools is greater than that of a FIB-SEM dualbeam tool (and a sputter coater). The cost of a single implanter is therefore an underestimate in this comparison.

In their presentation of the PL results, the authors avoid making comparative statements about the PL *intensities* measured from regions implanted in the same experimental conditions across thin films with different compositions. This is a wise approach, because quantitative estimations of PL intensities across different measurement runs are always prone to different instrumental artifacts (different focusing conditions, spectral efficiencies of the spectrometer, laser excitation powers, etc.). Nonetheless, it would be interesting (if at all possible) to provide (either in the main text or in the supplementary information) some *cautious* evaluations of the creation yields of different types of defects, possibly using spectral features such as the first-order Raman line of diamond as a normalization factor.

Authors: As requested, we added a suitably (ie, cautiously) phrased section on rough estimates to the supplementary information of the revised manuscript (See Supplementary Information 3). We note, however, that the most important qualitative result we have on PL intensities is that we can control the creation of most emitter species down to single defect level (as proven by second order autocorrelation measurements. $g^{(2)}(0) < 0.5$).

"specified in Figure 2a as the number of Xe+ ions per spot exposure": it would be much more informative (and appropriate) to report the indicated fluence values as numbers of ions per unit surface area, i.e. by taking into account the estimated diameter of the irradiated spot. Otherwise, a sentence like "The threshold for observable emission [...] is on the order of 600, 1500, 3000, and 3000 xenon ions" is rather meaningless and cannot be generalized.

Authors: We included the fluence estimates in the revised text, and clarified the "threshold" discussion as is it explained below.

"A minimum number of primary Xe+ ions is required to fabricate color centres": The authors should make an effort to explain (at least tentatively) this peculiar threshold-like behavior.

Authors: We thank the Reviewer for pointing this out — our wording was incorrect. There is no threshold, but rather the process is stochastic and hence an average number of ions is needed to generate a single color centre. We re-phrased the text accordingly.

*The observation of the emission above a particular $Xe^+$ fluence is related to the stochastic nature of the activation process, further discussed in the Supplementary Information 3.*

"Lateral precision was determined using PL maps of spot-arrays of low dose GeV ensembles. [...]": as correctly explained by the authors, the reported procedure results in a (44 ± 4) nm of the *accuracy* in the positioning of the bright spot with respect to its nominal planned location, which does not correspond to the *size* of the spot itself. In this sense, it is not clear why it should be comparable with the 30 nm ion beam diameter. The FIB spot size should be more appropriately related with the size of the bright spot, which is of course hard to measure (but very interesting to know!) unless super-resolution methods are used. Furtremore, the spot size is relevant (as much as its positiong accuracy) for integration in 2D photonic crystal cavities. The author should provide clarifications on these points.

Authors: We agree that both the positional accuracy and the beam spot size are important. We therefore clarified and expanded this discussion. We included an experimental analysis of the ion beam size. We also discuss the role of straggle in the precision with which a dopant is implanted in the target. We simulated the radial range and straggle for the case of Ge doping of diamond by an ideal (zero diameter) 30 keV $Xe^+$ beam and show that the resulting contribution (for this particular case) is ~6 nm, and places a fundamental limit on accuracy, precision and resolution under these irradiation conditions. Hence, the ion beam size in our PFIB instrument is the dominant factor in determining resolution. However, it is currently a mere technological limit due to aberrations rather than a fundamental one, and it is expected to improve with time.

We note that the size of the bright spots in PL maps is a measure of the confocal PL system resolution, rather than that of the implantation technique.

"We assume that ion channeling is the main factor responsible for the increased luminescence at this depth.": this tentative explanation is questionable, because it is hard to believe that a non-negligible fraction of the knocked-on atoms can preserve the very well defined directionality which is necessary for channeling to occur in the diamond crystal. This however is only a qualitative argument, that needs to be more carefully assessed by trying to quantitatively estimate how "residual" is the "residual PL signal [...] still present at a depth of (8 ± 2) nm", i.e. what is the fraction of channeling ions in the authors' estimations, and whether such fraction is realistically comparing with the modelling of the channeling process.

Authors: We agree. To address this point and other comments made by the Referees, we substantially expanded the modeling component of this work. Depth profiles simulated using a greater number of ions and vastly improved statistics show that the tail of the simulated depth distribution shown in the revised Figure 3b is, in fact, consistent with the value of (8 ± 2) nm measured in our experiments. Channelling is therefore not needed to explain the experimental data.

We also performed an analysis of the Ge momentum distribution at the film/diamond interface. It revealed that Ge does indeed lack the strong forward-directionality of the incident Xe, suggesting that channeling of Ge is only of minor significance.

"in accord with previously reported values—1.5 ns, 4.7 ns, 3.9 ns, 3.5 ns": the authors should provide the uncertainties associated to these estimated values.

Authors: We added the uncertainties to these values.

The bibliographic references suitably address the current state of the art in the field, and only some minor remarks are necessary.

References [13] and [16] could be suitably integrated with the first report on Sn-related color centers in diamond, i.e. [S. Ditalia Tchernij et al., ACS Photonics 4 (10), 2580-2586 (2017)].

Authors: We added the relevant reference.

The sentence "color centres in diamond, particularly the group IV-related [...] are currently front runners [...]" could be usefully integrated with a citation to the (elsewhere mentioned) ref. [10].

Authors: We added the relevant citation at this point in the text.

"We did not observe anti-bunching from the fabricated SiV color centres, most likely due to their very low quantum efficiency.": this observation should be complemented with a suitable bibliographic citation, specifically pointing at SiV color centers created by ion irradiation.

Authors: We added a reference to this statement.

"This value is better than the required threshold for engineering of quantum emitters within photonic devices, e.g. in the high field of a 2D photonic crystal cavity": this sentence would benefit from a suitable bibliographic reference

Authors: We added a reference to this statement.

From a formal point of view, the article is clearly presented, with the exception of very few points:

"one of the most fundamental principles of physics -": this expression sounds a bit bombastic and could be avoided without affecting the main message of the manuscript;

Authors: We rephrased the text in the abstract with *"...by exploiting momentum transfer from accelerated ions ..."*.

- red and purple points in Fig. 1: the color distinction in the figure is not very clear, as only purple points seem to be represented in the schematics;

Authors: We increased the color contrast to improve visibility.

- the portion of the periodic table reported in fig. 1b seems largely unnecessary to the stated scope of (cit.) "highlighting the versatility of this technique";

Authors: We agree and have removed the periodic table.

- I would recommend replacing the term "dose" with the more technically appropriate term "fluence";

Authors: We replaced the term in the revised version of the manuscript.

- "color centres": "color" is US English, while "centre" is UK English.

Authors: We changed all text to UK English.

Reviewer #2 (Remarks to the Author):
Review of: "Knock-on doping": A universal method to direct-write dopants in a solid state By Froch et al. March 2020
This paper presents a series of interesting results using the well-know technique of ion beam mixing. See for example the review by B.R. Appleton, Ion Implantation and Beam Processing, 1984 or indeed reference 30 in the present paper. The authors present some new applications for the method that would not have been contemplated in the 1980's. While it is worth bringing the technique to a new generation of researchers and employing new technologies to produce the ions, the authors do not make a compelling case for "significance" as required in the terms of reference of the journal.

Authors: We clarified and expanded our case for significance. In particular:

1. Our method features high lateral resolution (without the need for mask-based lithography), shallow implantation depth and extremely high dopant species flexibility. Whilst existing instruments can achieve certain aspects of our method individually, none features this combination of capabilities at once, which is the key to our claims of novelty and significance.
2. To strengthen our claims of "dopant species flexibility/versatility", we added new experimental data to the manuscript demonstrating that dopants from vapor-phase precursors can be delivered to the target and implanted by the $Xe^+$ ion beam. We used this approach to fabricate nitrogen-vacancy (NV) color centres by $Xe^+$ ion beam irradiation of diamond in the presence of gas-phase $N_2$. These results show that our technique is applicable not only to solid-state precursors supplied in the form of a thin film, but also to gases, as well as liquids and solids whose vapors can be delivered to the target using gas injection methods that are available on most commercially-available FIB instruments (as is explained in the revised manuscript and supporting references). The technique is thus readily applicable to direct-write implantation of, essentially, all isotopically-stable elements.

3. The new experiment we conducted with nitrogen (see 2), which we included in the revised manuscript, also shows that the thin film deposition step is not essential in our technique. We achieved <u>direct-write doping with high spatial resolution</u> whilst eliminating the need for all pre- and post-processing steps associated with conventional lithographic methods. This capability does not exist in the ion beam mixing literature.

4. Perhaps most importantly, our claims of a significant advance in the evolution of this technology are evidenced by the applications that we present in the manuscript. In contrast, the majority of the ion beam mixing literature is focused on the physics of ion-solid interactions. However, despite the long history of this field, no "killer applications" have been identified, and demonstrations of applicability are limited to effects such as changes in electrical doping of semiconductors. Conversely, in the revised manuscript, we demonstrate two applications of direct relevance to timely, high-impact research fields. **1)** The fabrication of single photon emitters with control down to the single defect level, and **2)** We added a second demonstration to the revised manuscript showing the fabrication of colour centres directly in the head of a single mode optical fibre. We selected the fibre because it is, objectively, a very difficult sample to process using conventional lithographic methods. Our choice showcases the applicability of our spatially-resolved implantation method to geometrically complex, 3D, non-planar targets.

To summarise, our technique represents a significant advance because it integrates all in one platform, a combination of features that typically require multiple instruments/methods. This can potentially give access to advanced applications which are unprecedented in the ion beam mixing field and are extremely difficult to achieve using any other implantation technique.

Title: Strictly speaking this is not a universal method. It is restricted to elements that can be deposited as a thin film. This covers much of the periodic table but will be problematic for some elements such as Li which is a dopant in diamond. Use of compound films will not address this problem because of co-doping from the other elements in the compound which cannot be avoided. It would be more appropriate to use the term "versatile" instead of universal.

Authors: We replaced "universal" with "versatile" as suggested by the Referee. We note, however, that the above-mentioned nitrogen implantation experiment demonstrates that our technique is, in principle, compatible with all elements that are isotopically stable. Species such as Li are challenging because they are susceptible to oxidation, but are not fundamentally incompatible with the method. For example, a Li micro-rod can be cleaned (by sputtering the oxide with a $Xe^+$ beam) and deposited onto the sample (also via sputtering by $Xe^+$ in a geometry in which the sputtered Li re-deposits on the target moved into place after the cleaning step), all in-situ, in high vacuum. Such a procedure would be in essence a form of Ion Beam Assisted Deposition, where instead of a small micro-rod a larger target is irradiated with a broad ion beam.[Martin, Philip J., et al. "Ion-beam-assisted deposition of thin films." *Applied Optics* 22.1 (1983): 178-184.]

Abstract line 25: The statement about the introduction of dopants through direct ion implantation remains a fundamental challenge is not correct. It is the method of choice for most of the IT semiconductor industry and is the essential for tailoring materials for numerous research applications.

Hence the word "challenge" is not appropriate given the method is both well understood and widely applied. The authors should reword this statement to make it clear what they mean.

Authors: We removed the statement about the introduction of dopants being a fundamental challenge and rephrased the abstract to better articulate the significance and novelty of our technique.

Abstract line 27: The author's method is in fact a form of ion implantation, with the implanted ions accelerated by forward recoil collisions ('momentum transfer" in the authors' words) with other ions. The authors should make a case for the advantages of their method over simply implanting the ions employing a low energy ion implanter instead.

Authors: We clarified the text to better explain the benefits of our technique, as we detailed above. We also included the above-mentioned results on implantation of vapour-phase dopants in the manuscript. Briefly, the key advantage of our method is that it enables direct-write, site-specific implantation of dopants from any solid film or surface-adsorbed gas molecule with high positional accuracy and resolution. This is particularly advantageous when the sample is a device that is objectively not easy to handle, such as the facet of an optical fibre (shown in the revised Fig 4a). Performing conventional lithography on it is challenging, but performing site-specific implantation using our technique is straightforward, as we demonstrate in the revised manuscript.

This application, whilst possible in principle, is extremely challenging using mask-based lithographic methods due to the need to perform lithography on a fibre head (i.e. a geometrically complex, small, high aspect ratio, non-planar substrate). Yet, it is easily achievable using our method due to the ability to image the target by an electron beam that is coincident with the ion beam.

For completeness, we also note that direct implantation of ions by highly focused ion beams (i.e. those focused to diameters of tens of nanometers) without the use of lithographic masks is problematic for a number of reasons. At present, there is no single focused ion beam implanter/technology that can accomodate a broad range of solid, liquid and gas sources that can be switched rapidly. Consequently, versatility is compromised and the spatial resolution of focused ion beams depends strongly on the ion species (particularly at low energies, due to the limitations of the ion focusing optics). Conversely, in our approach, a well-focused inert ion beam is used and the dopant species can be switched rapidly and easily.

Hence, our technique is highly advantageous over "simply implanting the ions employing a low energy ion implanter" — both the focused beam variety, as well as a broad beam combined with conventional lithographic techniques.

Abstract line 31: Justification for the claim of sub-50 nm lateral precision should be provided given the limitations of straggling of both the incident ions and the forward recoils would be substantial, even at these low energies. Also of interest is the depth precision where the low energy ions will get captured into the crystal channels and likely end up deep below the surface as a result.

Authors: We expanded the modeling component of this work and addressed these points explicitly. In the case of Ge implantation into diamond by a 30 keV $Xe^+$ beam, lateral straggle degrades resolution by ~6 nm (Supplementary Information 11). The lateral precision achieved by existing plasma FIB instruments is therefore limited by the ion beam diameter, as determined by the virtual source size and the focusing optics. We added an experimental analysis of the ion beam diameter to the revised manuscript (Supplementary Information 11).

Regarding depth-precision, we re-calculated the simulated Ge depth distributionusing a far greater number of ions to improve statistics. This improved simulation (performed in the absence of channelling) yields the depth profile shown in Figure 3b, which is in good agreement with our measured value of (8 ± 2) nm, and indicates that channelling effects are negligible. To confirm this, we analysed the momentum distribution of Ge during implantation, and showed that Ge does not possess the strong forward-directionality of the incident Xe species (see Supplementary Information 13).

Abstract line 32: The term "powerful" is not appropriate in this context for the same reasons as I object to "universal". Once again, I believe the authors mean "versatile" instead.

Authors: We rephrased the text and adopted the suggestion from the Reviewer: "versatile"

Page 2 line 51: The comment about ion implanters being expensive, etc, is a rash generalisation. Some laboratories employ simple desk-top machines which are adequate for low energy implants of colour centres in diamond and these systems are definitely not "expensive" or "not readily available". This statement should be deleted.

Authors: We rephrased the text to better explain the cost-benefit of our method. It features the combination of high precision site specificity (without the need for mask-based lithography), shallow implantation depth and extremely high dopant species flexibility. The cost of an equivalent laboratory based on conventional implanters and lithographic tools is far greater than that of a FIB-SEM dualbeam tool (and a sputter coater). A desktop implanter does not feature the totality of these capabilities, and is therefore not a useful reference for a cost comparison.

Page 2 line 56: Again, the statement about 10 nm implant depth being challenging is misleading given the large number of applications for such implants in the literature. It should be deleted or clarified.

Authors: We modified the text to clarify the context. Whilst existing instruments can achieve certain aspects of our method in isolation, none feature the combination of direct-write (mask free) lateral resolution, shallow implantation depth and species versatility/universality. The combination of these is key to our claims of novelty and significance, and the reason why we claim that our approach has broad appeal and applicability.

Page 2 line 58: The statement about "dose series" is unclear. Most implanters can implant a number of different ion species in the same sample without breaking vacuum. This is certainly not "practically impossible" as claimed here but is instead routine. This statement should be deleted or reworded to clarify the intent.

Authors: We disagree with the Reviewer on this point. A single capability (such as the implantation of multiple ion species without breaking vacuum) is not a suitable point of reference for this comparison. The ability to implant patterns of different doses and essentially any species and with sub-micron resolution without breaking vacuum is unique to our technique — even if the patterns are as simple as series of adjacent rectangles fabricated vs fluences. Conventional lithographic methods require multiple masking steps to achieve this, and no single focused ion beam technique features the species versatility of our method. We clarified the manuscript to ensure it is clear that this is the intended meaning of the text.

Page 2 line 61: The statement "cost-effective" is qualitative and should be deleted because it is presented without justification. Reference 30 is to a paper on ion beam mixing from the 1985 by one the then leading groups in the field. The authors must justify what is new about their method and how it is a major advance from the previous work in the field.

Authors: We revised the text, as explained in the responses above addressing our claims of cost-benefit, as well as novelty and significance of our work.

Caption for Figure 1: The use of an electron beam to image implant sites for colour centres in diamond will likely introduce vacancies that can influence the formation rate and type of colour centres formed as a result. The electron irradiation is also likely to change the charge state of N-V colour centres from the neutral to the negatively charged state leading to complexities in reproducing results.

Authors: The initial submission was not concerned with NV centres. Nonetheless, we disagree in general with this criticisms for the following reasons:

- The threshold for vacancy formation by electrons in diamond via knock-on is >80 keV, which is much higher than that employed in SEM (here, we used 2 keV). [Koike, J., *Applied physics letters* 60.12 (1992): 1450-1452.]
- Electron damage generation rates through other mechanisms are negligible compared to that of ion bombardment/implantation.
- The colour centres are not formed during the initial implantation process but in the subsequent annealing step which induces vacancy diffusion and centre formation. Consequently, the electron imaging step has no effect on defect charge states (which can, in fact, be manipulated using a range of reversible surface processing and biasing methods).

Page 6 line 182: Suggest changing "GeV" to Ge-V" to reduce ambiguity with more common uses of "GeV".

Authors: The meaning of GeV, i.e. germanium-vacancy centre, is defined in the text. It is—a perhaps unfortunate—initialism, but one that has been deliberately chosen and employed extensively throughout the entirety of the relevant literature. On this basis we prefer to adopt it here for consistency.

Page 7 lines 186-189: These details are elementary and should be discarded, or prefaced with a statement about the problem being solved with the method.

Authors: We agree with the Reviewer and moved this text to the Supplementary Information 10.

Page 7 line 191: A resolution of 44 nm is deduced from the distribution of colour centres. This figure should be compared to a convolution of both the beam spot size and the likely straggling from the forward recoils.

Authors: In the revised manuscript, we analyzed the role of radial range and straggle explicitly and showed that it is insignificant under the typical conditions used in our work (i.e., ~6 nm in the case of Ge implantation into diamond by a delta-function (zero diameter) 30 keV $Xe^+$ beam). The beam spot size of current-generation plasma FIB instruments is limited technologically and expected to improve in the future (see also the following response). We clarified this in the text.

Page 7 line 196: It is not true that plasma ion sources have an inherently larger beam pot with several manufacturers offering high resolution probes. The beam spot size also depends on the beam current so this factor also needs to be considered when making comparisons between different systems.

Authors: We agree that beam current affects aberrations associated with ion beam focusing. However, the beam spot size is ultimately limited by the virtual size of the source, which is relatively large in the case of the plasma ion source employed in our experiments. Other technologies such as a Gas Field Ionization source (GFIS) feature a smaller virtual source size and hence a smaller beam spot size. We revised the text to clarify this point as following.

*"... It should be noted that, in the case of the plasma FIB used here, the beam diameter is limited by the virtual source size and the ion beam focusing optics. It can be improved in a number of ways, such as utilisation of a Gas Field Ionisation Source…"*

Page 7 line202: The term "cavity hotspot" is not defined.

Authors: We rephrased this part to *"...the high field region of a photonic resonator…"*

Page 7 line 204-5: As the comment above, SEM irradiation can change the nature of colour centres so this statement should be deleted.

Authors: We disagree with this criticism as it pertains to colour centres in diamond, as explained above.

Page 7 line 208: The term "peculiar" is not appropriate in this context because the method of ion beam mixing is already well understood and widely studied in the 1980s as pointed out above.

Authors: We removed the term "peculiar" from the text.

Page 7 line 210: Colour centres in diamond created within 1 nm of the surface are likely very sensitive to the nature of the surface termination. It is not possible to assess the authors' claims here without knowing this essential information.

Authors: We clarified the text. See also the below response relating to "page 8, line 229".

Page 7: In the discussion of the technique the effect of forward recoils being captured into crystal channels and ending up deep in the crystal is ignored. Yet this is likely to be an important feature of their method, especially for light ions. Likewise the residual lattice damage from both the primary and forward recoils should also be accounted for in a discussion of the results.

Authors: We clarified the discussion of Ge channelling and ion-induced damage. We added a theoretical analysis of the momentum distribution of the Ge dopants, which shows that they do not possess the strong directionality of the incident Xe ions (Supplementary Information 13). This suggests that Ge channelling is insignificant, consistent with our range simulations and measurement of the Ge depth-distribution. We have expanded the discussion on the damage and emphasise that further treatment may be required.

Page 8 line 229: This statement is not clear. Also the proximity of the surface and also surface charge traps is likely to have a significant effect on the luminescence efficiency.

Authors: We rephrased this sentence to improve clarity. We note that this result/claim is not controversial as color centers have been observed in nanodiamonds with a diameter of ~1.6 nm (ie: a maximum depth of 0.8 nm below the surface) [Vlasov, Igor I., et al. "Molecular-sized fluorescent nanodiamonds." *N_xxxxxx* 9.1 (2014): 54-58.]

Page 8 line 236: There is a missing symbol after "x". Similarly, further occurrences below.

Authors: We thank the Reviewer for noting this and corrected the text accordingly.

Page 8 line 239: Here there is speculation about the role of ion channeling, but it is not clear if such dopants can be activated.

Authors: We revised the discussion of channelling, as is detailed above.

Page 8 Figure 3 caption: The nature of the dopants responsible for the results shown in this figure should be repeated in the figure caption itself. It is not clear from the text if Ge doping applies to all sub-figures.

Authors: We clarified this in the revised manuscript.

Page 9 lines 262-3: The description of the results as a "major breakthrough" is unjustified hype and detracts from the paper. It should be justified or deleted.

Authors: We removed this statement from the text.

Page 9 line 268: It is not clear how the method has advantages over just implanting the elements directly instead of employing the forward recoils. In some limited circumstances there may be operational efficiencies if the appropriate apparatus is readily available, but this will not always be the case.

Authors: We clarified the benefits of our approach, as it is detailed in our responses above. We also added new results showing knock-on implantation of gas-phase dopants in order to support our claim of species-versatility, and a demonstration of colour centre implantation directly into the head of a fibre to demonstrate the immediate applicability of the technique to challenging systems/devices.

Page 10 line 292: The statement about the irradiations performed with a day or less should be expanded to explain why this is an issue and what problems they are overcoming.

Authors: This statement was intended merely to explain the procedure. We have removed it from the revised manuscript.

In conclusion, the authors report a series of experiments to use forward recoil ions to inject near surface colour centres into diamond. The paper needs to justify the significance of the result to a greater extent to meet the publication criteria for N_xxxxxx.

Authors: Please see the above response to the first comment made by Reviewer 2.

Reviewer #3 (Remarks to the Author):
The manuscript by J.E. Fröch et al. entitled ""Knock-on doping": A universal method to direct write dopants in a solid state host" is indeed a very interesting study, which might be worth publishing in N_xxxxxx; however, it is written in a lurid manner containing many false statements motivating their own work. It might be that one has to write manuscripts in this way to overcome the editorial check, but this behavior discredits science. Thus, such a manuscript is not acceptable in any journal.

Authors: We thank the Reviewer for the remark that our work is very interesting and might be worth publishing in N_xxxxxx. We revised the manuscript and added additional experimental and modelling results in order to address all the concerns raised.

In detail, comments and questions in the order of appearance:

(1) Title. The authors introduce the term "Knock-on doping" in order to give the technique a new flavor. However, the authors themselves know [see references 29,30] that this technique is very old and has been already used for decades. Therefore, in order not to confuse the science community, the authors need to use the correct terminology "Recoil implantation" not only in the title, but throughout the whole manuscript. One might here also cite more relevant literature, such as [Surface Science 57 (1976) 143; J. Applied Phys 50 (1979) 7261; NIM B 7-8 (1985) 645, etc. ]

Authors: We agree that "recoil implantation" as well as the term "ion beam mixing" (flagged by Referee 2) are used in the literature. We named both explicitly in the revised introduction. We do, however, believe that the use of "knock-on" is justified specifically to give a new flavour to our approach, given the distinguishing features that we explain below and in our detailed first reply to Referee 2 above. Given the editorial nature of this request, we are however happy to let the editor rule on this point.

(2) Line 51 – A new ion implanter costs about 1-1,5 M€ (ask for a quotation from High Voltage Eng.). The here used Xe-FIB is much more expensive (typically > 2 M€) and even harder to maintain.

Therefore, to motivate the study being "cost-effective" in comparison with conventional ion implantation is absolutely false!

Authors: We deemphasized and clarified our claimed cost benefit. We believe that it is correct given the species-versatility of our method and the ability to direct-write dopants with high spatial resolution. Whilst existing implanters can achieve certain aspects of our method in isolation, none feature the same combination of direct-write (mask-free) lateral resolution, shallow implantation depth and species versatility — which we now demonstrated further in the revised manuscript by showing knock-on doping of N from a gas-phase $N_2$ into diamond to form nitrogen-vacancy (NV) colour centres (further details are provided below).

It is the totality of the benefits of our method that is key to our discussions of cost, novelty and significance. The cost of an equivalent lab that allows direct-write implantation with site-specificity that is precise to tens of nanometers using essentially any dopant species is far greater than that of a single FIB-SEM dualbeam tool (and a sputter coater). The cost of a single conventional implanter is not a suitable point of reference. We revised the manuscript to clarify the cost benefit of our technique.

The authors should have also in mind that about 20 ion implantation steps are necessary to produce any processor chip at Intel or AMD using conventional ion implanters, on the other hand, computers / mobile phones etc. are today very cheap.

Authors: The extremely low cost of consumer electronics is not relevant because it is dictated by the economics of mass-production (eg: over a billion smartphones are manufactured annually) rather than that of the equipment used to manufacture them. We also note that the ion implanters (as well as FIBs and other large pieces of equipment) used in Intel/AMD fabrication plants are in fact often very inflexible — they are designed to perform tasks that are very complex and impressive with extreme reliability and reproducibility, but this is often achieved at the cost of low flexibility, and any deviation from highly specific tasks is often challenging. For example, some high-end FIBs made specifically for fab plants by Applied Materials require a service engineer call-out to change the ion beam energy — a task that is performed routinely by students with a single mouse click using FIBs designed for research (rather than manufacturing), such as the one used in our work.

(3) Line 53 – Why are conventional implanters not ready for end-users? Operating an implanter is as easy / complicated as operating an SEM/FIB. Check out for facilities, which run both techniques.

Authors: We removed this claim and clarified that we were referring to the combination of features offered by our technique: direct-write mask-free implantation with high spatial resolution, and species versatility/universality. The latter is highlighted further by new experimental data that we added to the manuscript to underscore the readiness, universality and versatility of the technique. In a set of new experiments (added in the manuscript), we show that dopants from vapor-phase precursors such as $N_2$ gas can indeed be implanted by the $Xe^+$ ion beam (see our response below to the referee's comment on $N_2$ compatibility). These new results show that our technique is applicable not only to solid-state precursors supplied in the form of a thin film, but also to gases, as well as liquids and solids whose vapors can be delivered to the target using gas injection methods that are available on most commercially-available FIB instruments (as it is explained in the revised manuscript and supporting references). The technique is therefore compatible with all isotopically-stable elements in the periodic table.

(4) Line 54 - Ion implantation can be even done with ion energies down to 20 eV (!) using conventional implanters. It has been demonstrated that even one monolayer graphene can be doped [Nano Letters 13 (2013) 4902]. Therefore, the statement of the authors is false that "ion implantation is generally applicable only for relatively deep implants". Today, shallow implantations at Intel and AMD into Si are typically done with 1-2 keV ions resulting into doping profiles less than 10 nm (check with TRIM).

Authors: We revised the text to clarify that we are referring to the combination of capabilities featured by our technique: shallow direct-write implantation with high spatial resolution and species universality. Conventional low energy implanters do not feature high spatial resolution — instead, they require the use of masking/lithography to achieve site specificity with a precision down to few tens of nanometers. We did, however, add the above reference to the revised introduction.

We also note that it is the combination of multiple capabilities that allows our technique to achieve high-impact applications, as substantiated by the 2 examples in the revised manuscript:

1. Site-specific direct-write recoil-doping of a range of colour centres with control down to the single emitter level.
2. An additional experiment we added to the manuscript demonstrating site-selective colour centre implantation into the head of an optical fibre. We chose an optical fibre as it is, objectively, a very difficult sample to process using conventional lithographic methods. We illustrate the applicability of our spatially-resolved implantation method to geometrically complex, 3D, non-planar targets.

(5) Line 57 – Co-implantation of elements and ion fluences series (not "dose" series!) can be also done easily by conventional implanters (google scholar term "co-implantation" gives >8.000 hits, "ion fluence range" gives >251.000 hits).

Authors: We revised the text to clarify the intended meaning. The ability to implant patterns of different fluences and essentially any species and with sub-micron resolution without breaking vacuum is unique to our technique — even if the patterns are as simple as adjacent rectangles fabricated vs fluences. Conventional lithographic methods require multiple masking steps to achieve this, and no single focused ion beam technique features the versatility of our method in terms of species selection.

Addendum to (1-5): I recommend that the authors visit/talk to their colleagues at ANU in Canberra, where conventional ion implantation is performed since decades, and get more correct knowledge and up-to-date information on ion implantation, before comparing it to recoil implantation by FIB. Furthermore, the authors should be aware that their technique does not allow for isotopic pure recoil implantations neither for gaseous species, such as nitrogen.

Authors: We added an extra result to the manuscript to demonstrate both the versatility of performing recoil implantation using a FIB-SEM dualbeam, and that it is in fact compatible with gas species such as nitrogen. Specifically, we showed that dopants from vapor-phase precursors can be delivered to the target (at room temperature) and implanted by the Xe$^+$ ion beam. We demonstrated this by utilizing a capillary-style gas injector; we used this approach to fabricate nitrogen-vacancy (NV) color centres by Xe$^+$ ion beam irradiation of diamond in the presence of gas-phase N$_2$. We remark again (see point 3), that these new results show the applicability of our technique not only to solid-state precursors in the form of a thin film, but also to gases, as well as liquids and solids whose vapors can be delivered to the target using gas injection. We added a detailed explanation in the revised manuscript and supporting references. We also note that the vapour-phase approach yields an additional benefit for direct-write, localized doping: film deposition and removal steps are not necessary.

We disagree that the technique is not isotope-compatible. Any element that is isotopically stable can be either deposited as a solid film, or delivered to a target using the vapour-phase variant of our technique (most major suppliers of research-grade chemicals and gases offer isotopically-pure species). We also note that any element that is not isotopically stable is not relevant to our claims since it would decay in the target after incorporation by any technique.

(6) Line 86 – What was the ion fluence for writing the logo?

Authors: We added the fluence (2.5×10$^{13}$ cm$^{-2}$) for the creation of the logo to the text.

(7) Figure 2 - (a) The authors use here the term "dose" while knowing that the correct term is "ion fluence", as used in the SI. Furthermore, the numbers in the image do not have a unit! I guess that these are amounts of Xe ions, as described in the text. However, this quantity does not help any scientist to reproduce the experiments, as the area needs to be known. Figure 2 (b) & Line 138 - Which ion fluences were used for these spectra?

Authors: We agree. We have added the fluence values to the relevant text and to Figure 2a.

(8) Line 130 – Why is the minimum numbers to create color centers so high? In principle, one single Xe ion should be enough to create one color center! Likely, the PL system is not sensitive enough.

Authors: We have rephrased the text and added further discussion about the creation yield (Supplementary Information 3) to clarify this point. In brief, the yield, i.e. the number of optically-active color centers created per ion implanted is less than unity. In fact, it can be as low as 0.01, like in the case of Si. Note that this is not a limitation of our technique but instead a consequence of the fact that the fluorescent defect complexes are formed stochastically during a post-implantation anneal.

The sensitivity of the photoluminescence (PL) confocal system is not an issue: in fact, we can detect individual, single-photon emitters (as shown by the second order autocorrelation functions, $g^{(2)}(0) < 0.5$, indicative of sub-Poissonian photo-emission statistics).

(9) Line 191 – The precision of 44 nm is nice, but the authors should compare this with existing technologies, such as implantation through a hole in an AFM tip, which yield into a precision of < 16 nm. [Small 6 (2010) 2177]

Authors: We revised our discussion of precision and resolution. Briefly, the most relevant parameter in our work is the positional accuracy with which a dopant can be implanted, which is underpinned by the ion beam spot size (we added an experimental analysis to the supporting information) and straggle (which plays a minor role; see below). We cited the paper suggested by the Referee. We note, however, that this paper does not report an implantation precision of 16 nm, but rather the ability to resolve emitters that were created—stochastically—within 16 nm of each other in a single implantation spot (i.e. the 16 nm relates to the spatial resolution of STED, the employed super-resolution fluorescence imaging technique).

(10) SI – section 2 – The authors use TRIM to calculate the recoil ion ranges. This is OK for the situation of low ion fluences, but not OK for high fluences, such as > 5x1014 cm-2. Here, the authors should use TRIDYN to account for dynamic changes while accumulating implantation, sputtering, ion beam mixing, etc. Furthermore, the statistics of the presented simulations is too low, and the authors should plot figure S3a on a log-scale.

Authors: We substantially expanded the modeling component of this work:

- Improved statistics yields good agreement between the tail of the simulated depth profiles (seen in Figure 3b) and the actual depth measured in our PL experiments, (8 ± 2) nm.
- We analyzed the contribution of lateral range and straggle to the lateral resolution and show that it amounts to ~6 nm for the case of Ge implantation into diamond by 30 keV $Xe^+$.
- We performed a detailed analysis of the momentum distribution of the Ge dopants at the film-diamond interface, which shows that they do not possess the strong directionality of the incident Xe ions. This suggests that Ge channelling is insignificant, consistent with our range simulations and measurement of the Ge depth-distribution
- We increased the statistics of the calculations, presented in Figure 3b, and plotted them on a log scale.

Instead of TRIDYN, we used our own code that allows us to implement and evaluate a broad range of mechanism, and data extraction/analysis algorithms. The Referee was likely contemplating the following effects, which we evaluated and dismiss because they are insignificant under our experimental conditions for these reasons:

1. At an ion current of 10 pA (relevant to our analysis of Ge depth and lateral distributions), the average time between the arrival of ions at the target is ~16,000 ps. On the other hand, the time scale for the ballistic atom collision processes is on the order of 1 ps. Hence, nonlinear processes arising from collision cascade overlaps can be neglected.
2. The (Ge) thin film is sputtered by the ion beam during implantation. Hence the thickness of the film at the end of the process is reduced and this will increase the Ge depth profile in diamond. However, under the conditions used to analyze Ge spatial distributions, this effect is negligible because the initial film thickness was ~15 nm and decreased by only ~2 nm.
3. The high fluences quoted by the referee (i.e., >5×$10^{14}$ $cm^{-2}$) were not used in the experiments that we used to evaluate spatial distributions of Ge in diamond (we clarified this point in the manuscript). Instead, fluences on the order of ~$10^{13}$ $cm^{-2}$ were used, and are more directly relevant to the application we focused on (single emitters), and also more insightful for establishing the ultimate resolution of the technique.

3rd June 2020

Your manuscript entitled "Knock-on doping: A versatile method to direct-write dopants in a solid state host" has now been seen by 3 referees, who also saw the earlier version of your manuscript. Please find their comments attached below. In the light of their advice, we have decided that we cannot offer to publish your manuscript in N_xxxxxx.

The reviewers appreciate the newly added information, data, and experiments and your response to the technical queries of the earlier round of review. And again, they emphasise the quality of your work. However, in their reports, reviewer#2, but also reviewer#3, raise again concerns regarding the advantages of your method in terms of widespread applicability, cost, or precision, over existing methods. As such potential advantages made us consider your work in the first place, the lasting criticism of the reviewers for us precludes publication of your work in N_xxxxxx.

In particular, reviewer#2 doesn't feel that a convincing case is made that the method has considerable advantages over ion implanter systems. She/he misses a convincing demonstration that your method can achieve functionalities or designs not accessible with alternative methods. In addition, reviewer#3 is concerned about the validity of your arguments regarding the costs of the method compared to alternatives. And while we understand that there is some support from reviewer#1 and reviewer#3 for publication, the arguments above and some more technical concerns of the reviewers are important enough for us to no longer consider your manuscript. Reviewers Comments:

Reviewer #1 (Comments for the Author):

The authors suitably addressed the remarks that were raised in the first review, therefore the paper is considered suitable for publication in its present form.

Reviewer #2 (Remarks to the Author):

Review of revised manuscript: Knock-on doping: A versatile method to direct write dopants in a solid state host, by Fröch et al.
The revised manuscript addresses most of the technical concerns raised by the referees with many new details of the experiments and results.
The unnecessary hyperbole has also been removed.
However, these essential changes have not addressed the main concern: the stated purpose of N_xxxxxx is to publish "… papers of the highest quality and significance". The manuscript is now of high quality but does not meet the significance criterion. This is because of the following reasons:

(1) As pointed out in several reviews, the authors have not made a strong case that their methods are significantly novel compared to the ion beam mixing work from some time ago.
(2) There is an unconvincing case made that the advanced and expensive instrumentation employed for the experiments is more convenient and accessible than standard and less complex ion implanter systems. In fact, a much simpler system has already been demonstrated, see "Spin measurements of NV centers coupled to a photonic crystal cavity", Jung et al., APL Photonics 4, 120803 (2019); https://doi.org/10.1063/1.5120120. In this work the authors employed an AFM nanostencil and a gas plasma source suitable for gas and solid state source material.
(3) The rejoinder claims "The fabrication of single photon emitters with control down to the single defect level" but this is not substantiated in the revised manuscript where no such control is demonstrated. This is because the method cannot be used to fabricate and control single emitters any better than just broad beam irradiation combined with post implant searches for isolated single emitters created randomly by the beam.
(4) The addition to the revised manuscript of examples of forward recoils from gas atoms in the vacuum system raises questions about the effect of unavoidable contamination of unwanted dopants introduced from residual gases in the vacuum system.
(5) The addition to the revised manuscript of the implantation of an optical fibre is not novel because there are already examples in the literature of the fabrication of Bragg gratings and other structures in fibres by ion implantation.
(6) The added uncertainties in the fabrication process arising from the co-implantation of the primary ions is likely to deter widespread take-up of the method. This means the method would probably not be viewed "Of extreme importance to scientists in the specific field" as required by the journal publication criteria.

Finally, the authors have presented a sophisticated series of experiments which are described carefully in great detail about a method of creating random numbers of colour centres in a nanoscale localised area. What is missing is a significant problem that the method can address that cannot be solved by methods already published in the literature which may include a visionary device architecture with revolutionary potential applications.

Reviewer #3 (Comments for the Author):

The authors revised the manuscript and addressed most of the issues raised by the three reviewers in a proper way. However, I absolutely cannot recommend publishing the present version, because the authors still ignore good scientific practice.

(1) In agreement with reviewer two, I pointed out that all single aspects of the presented technology are not new. Reviewer #2 even asked for the significance. In contrast, I do see the significance in the combination of all applied aspects, which I mentioned in my previous report, and I do appreciate the two new applications (Eu3+ in fibers, nitrogen) the authors presented in the revised manuscript. However, the flavor of the writing and presentation is still not suitable for any scientific journal. The authors like to give an old technique a new flavor by renaming the process (answer letter, page 12 mid). This behavior discredits science and furthermore polarizes the science community – especially people from the ion beam community will be upset. There is no need for this article nor for N_xxxxxx to do tabloid journalism! I strongly recommend to replace all lurid phrases, such as "knock-on",

"new", … with appropriated scientific terms throughout the
manuscript, as I suggested already in my first review report. Including the title.

(2) Next is the discussion on the "costs", which was also full in line with reviewer 2. I agree that an implanter in combination with photolithography system (which btw can indeed make today structures on the 10 nm level) is more expensive than a Xe FIB/SEM & sputter coater. Sure, these are the investments. However, the authors compare apples and oranges! I like to see how cheap it will be, when the author will structure a 450 mm diameter wafer with their approach. I guess they will need a decade. On the other side, multiple large-scale wafers can easily be implanted within minutes by the conventional route. Maybe the authors need a lecture on full-cost pricing at a business school, and then they will understand the extreme low costs of conventionally implanted consumer electronics with 20+ photolithography and implantation steps to make multiple element and (even 3D) concentration gradient structures. The key point: WHY is it necessary to discuss costs in a physics article?
There is no need for this, therefore I recommend fully removing this aspect throughout the manuscript, and I give the advice to the authors not to use such an argument at any oral presentation. This again would only polarize the community.

(3) Answer letter, page 6, comment to reviewer 2. The sentence "…"no killer applications" have been identified ... and … are limited to effects such as changes in electrical doping of semiconductors" is clear cockiness of the authors on their field. On one side, today there is no single killer application for color centers in diamond, which I can buy in a store, and on the other hand, consumer electronics – including the mobile phones of the authors – rely on the changes in electrical doping of semiconductors by conventional implantation! Again – why do the authors need to polarize the community? There is no need for such a behavior.

(4) Lines 56 – 58. The sentence "They (… standard implanters …) are also generally ineffective for the implantation of foreign species into atomically-thin materials such as graphene or transition metal dichalcogenides, … " is wrong. The authors should read the reference in detail. True is that any element into any 2D-material can be implanted using standard implanters, because they provide ion energies as low as 20 eV! Up to now there is no single FIB-system, which can provide such low ion energies!

(5) Even though I appreciate the experiments with nitrogen – supplementray information 2 – these experiments are not convincing at all. Figure S3(b) does not show an PL-intensity increase as a function of ion fluence for the field 1 to 4, and even worse: the PL-yield seems here to be comparable to the not irradiated area between fields 2 and 6, 3 and 7, 4 and 8. Only field 5 shows a clear signal.

10/06/2020
Dear Dr ___,
Before we address in details the referee's comments, it is important to look back at their comments. I am sure you would agree that referee #3 is recommending acceptance and the sole reservation is with the style. In fact most of the comments refer to what we've written in the response letter and not in the actual manuscript, **AND** the referee identified clearly the significance of our method and its new applications. On the other hand, we are confused and disappointed with the biased tone and

**incorrect statements** of reviewer #2. The referee clearly "caved" on their original assessment which included a long list of questions, which we **have fully (!)** addressed, and now bluntly citing the journal guidelines and saying "this work does not meet those" without citing any prior literature. If I am a reviewer for you, Ben, would you really consider such an empty claim of "it's not new"?

There are only a handful of examples where an old technology "unexpectedly" produces "state-of-the art" science. This on its own is what pushes the scientific boundaries, promotes innovation and triggers the non-conventional line of thought. This is what inspires the new generation of scientists, and this is what our research is about. The current work is not aimed at replacing an ion implanter. One should be naïve to think that way. Same as graphene did not replace the transistor and the NV centre did not replace the SQUID. Yet, it is the unlimited unexplored potential of a method that exists in the vast majority of universities and can indeed be exploited to study cutting edge science — in a way that no one envisioned before (as even referee #2 acknowledged in the first round!). Graphene is not amazing because you can measure the Hall effect at room temperature. It is amazing because you only need a scotch tape to do so, rather than a highly sophisticated GaAs MBE chamber. Likewise, almost every university has a FIB these days, but only a few countries have, as a matter of fact, a commercial ion implanter.

We did not claim that our results will "achieve functionalities or designs not accessible with alternative methods" — as stated in your letter. Ironically, however, throughout the revision we realised we can make this claim. And we proved it experimentally by producing luminescence from emitters implanted right at the facet of an optical fibre; something that our technology can achieve, uniquely. I dare you or referee #2 find a reference that does this with a conventional implanter. But more importantly, we are the first to identify that an "old" technology from the 80s can be used today, and generate cutting edge results in quantum technologies, nanophotonics and nanotechnology. Don't you think this fact alone is inspiring? Combining with a straight acceptance recommendation of reviewer #1, and the stylistic recommendation from reviewer #3, which we took on board, we strongly believe our work is suitable for N_xxxxxx.

Finally, we would be happy for the paper to be revisited again by reviewer #3, or even sent to a 4[th] referee, but as you would see below, we would prefer if you wouldn't send it back to reviewer #2.

Igor Aharonovich

Reviewer #2 (Remarks to the Author):

The revised manuscript addresses most of the technical concerns raised by the referees with many new details of the experiments and results. The unnecessary hyperbole has also been removed.

However, these essential changes have not addressed the main concern: the stated purpose of N_xxxxxx is to publish "… papers of the highest quality and significance". The manuscript is now of high quality but does not meet the significance criterion. This is because of the following reasons:

Before we address these comments in details, it is critical to note that the reviewer throws out controversial statements, which are simply wrong and are not substantiated by any reference.

(1) As pointed out in several reviews, the authors have not made a strong case that their methods are significantly novel compared to the ion beam mixing work from some time ago.

As was explained in the first revision, the works on ion mixing were a proof of principle study. This method was hidden in the literature and not used successfully for **any** application. We identified it, and found critical applications that work. This is in fact a major, radical advance over what has been done.

More specifically, ion beam mixing/recoil implantation was mostly studied with a focus on mixing at the interface of adjacent layers. In part this was done out of scientific interest with no application in mind and mainly to attempt at studying physical effects occurring during this process (1-3). Some papers mention the advantage of increasing the surface adhesion and to utilise this characteristic e.g. for chemical processing under harsh conditions (4) or wafer bonding (5). One recent report mentioned the utilisation of a SiC layer fabricated by ion beam mixing (6). Earlier attempts tried to look into the realisation of electrical doping, or metal contacts to reduce gate/source/train resistances but that was **NOT** actually demonstrated (7-11). Most relevant, there was one paper reporting *an attempt* of implanting Er into Si, which ***however did not show a spectrum of the implanted material*** (12).

Objectively, we succeeded in demonstrating an effect that was pursued by the community for 40 years and abandoned due to major technical difficulties which we have overcome with an elegant manner. This is, in our view, the definition of "significantly novel".

(2) There is an unconvincing case made that the advanced and expensive instrumentation employed for the experiments is more convenient and accessible than standard and less complex ion implanter systems. In fact, a much simpler system has already been demonstrated, see "Spin measurements of NV centers coupled to a photonic crystal cavity", Jung et al., APL Photonics 4, 120803 (2019); https://doi.org/10.1063/1.5120120. In this work the authors employed an AFM nanostencil and a gas plasma source suitable for gas and solid state source material.

We appreciate the reference suggestion, and acknowledge the importance of that work but it has little to no relevance with regard to our work. We really are not sure how an AFM integrated system with the FIB is simpler than just a FIB. This is objectively, nonsense.

(3) The rejoinder claims "The fabrication of single photon emitters with control down to the single defect level" but this is not substantiated in the revised manuscript where no such control is demonstrated. This is because the method cannot be used to fabricate and control single emitters any better than just broad beam irradiation combined with post implant searches for isolated single emitters created randomly by the beam.

Once again, we are confused by this comment. Single emitter arrays (!) are shown in figure 2 — all autocorrelation functions (insets) prove single emitters. how can this be missed? There is no need in any post implant "search" since we are making an array.

(4) The addition to the revised manuscript of examples of forward recoils from gas atoms in the vacuum system raises questions about the effect of unavoidable contamination of unwanted dopants introduced from residual gases in the vacuum system.

There is no evidence of residual gas contamination in our results. The spectrum from a reference irradiation performed in the absence of N2 that we provided shows no NV formation. Moreover, contamination from gases is not a problem in general – as we explained in revision 1, these gas injection methods are used routinely in these instruments – without problems. Once again, the referee makes claims that are strong, controversial and wrong without backing them up.

Nevertheless, as referee #3 also commented on that, we repeated this measurement and showed a better set of results with increasing intensity from the formed NV centres, as requested by referee #3.

(5) The addition to the revised manuscript of the implantation of an optical fibre is not novel because there are already examples in the literature of the fabrication of Bragg gratings and other structures in fibres by ion implantation.

This is an incorrect statement. While fabrication of Bragg grating is done using ion implanters, it is NOT creation of optically active defects. Our process is not about making structures; its about generating fluorescence. Surely the referee is aware of that. For the referee's benefit, we include ref [13-15] where this work is performed. The referee may find it interesting, that in fact, in all those works, the implantation is done through the cladding (i.e. the long side of the fibre) rather than the "top facet" as done in our work.
We are disappointed with this uninformed comment from the referee.

(6) The added uncertainties in the fabrication process arising from the co-implantation of the primary ions is likely to deter widespread take-up of the method. This means the method would probably not be viewed "Of extreme importance to scientists in the specific field" as required by the journal publication criteria.
Once again — for us to respond appropriately, we need to understand what uncertainties does the reviewer refer to? In fact, we added in the revised version a very detailed modelling analysis of the depth profile of the ions, exactly for that reason and as the referee originally requested (comment about sub 50 nm, first revision), and the reviewer rather conveniently chose to ignore those. Quoting the guidelines and saying, bluntly "it's not going to be adapted" is a highly subjective and inappropriate comment.
Finally, the authors have presented a sophisticated series of experiments which are described carefully in great detail about a method of creating random numbers of colour centres in a nanoscale localised area. What is missing is a significant problem that the method can address that cannot be solved by methods already published in the literature which may include a visionary device architecture with revolutionary potential applications.

The vision and the potential applications are written in a scientific and appropriate manner in the abstract, introduction and the conclusions of the manuscript. In fact, this same referee already commended that *"The authors present some new applications for the method that would not have been contemplated in the 1980's."* The same was also explained in the original response letter, and in light of the unsubstantiated comments from the referee, we doubt any answer will ever be satisfactory to them.
Reviewer #3 (Comments for the Author):
We first would like to apologise that our language seemed to upset the reviewer. It was not the intention, and most of the argument were indeed kept to the response letter. We now corrected it all throughout the manuscript as detailed below.
The authors revised the manuscript and addressed most of the issues raised by the three reviewers in a proper way. However, I absolutely cannot recommend publishing the present version, because the authors still ignore good scientific practice.
We thank the reviewer for acknowledging that the issues were addressed in a proper way and below, would like to ensure we also corrected the style to meet the adequate scientific practice.
(1) In agreement with reviewer two, I pointed out that all single aspects of the presented technology are not new. Reviewer #2 even asked for the significance. In contrast, I do see the significance in the combination of all applied aspects, which I mentioned in my previous report, and I do appreciate the two new applications (Eu3+ in fibers, nitrogen) the authors presented in the revised manuscript. However, the flavor of the writing and presentation is still not suitable for any scientific journal. The authors like to give an old technique a new flavor by renaming the process (answer letter, page 12 mid). This behavior discredits science and furthermore polarizes the science community – especially people from the ion beam community will be upset. There is no need for this article nor for N_xxxxxx to do tabloid journalism! I strongly recommend to replace all lurid phrases, such as "knock-on", "new", … with appropriated scientific terms throughout the manuscript, as I suggested already in my first review report.

We are glad the referee acknowledges the significance of the paper and clearly understood the extra results. We take on board the comments about the wording, and replaced knock-on with recoil implantation. The word *"new"* was in fact never used throughout the manuscript.

(2) Next is the discussion on the "costs", which was also full in line with reviewer 2. I agree that an implanter in combination with photolithography system (which btw can indeed make today structures on the 10 nm level) is more expensive than a Xe FIB/SEM & sputter coater. Sure, these are the investments. However, the authors compare apples and oranges! I like to see how cheap it will be, when the author will structure a 450 mm diameter wafer with their approach. I guess they will need a decade. On the other side, multiple large-scale wafers can easily be implanted within minutes by the conventional route. Maybe the authors need a lecture on full-cost pricing at a business school, and then they will understand the extreme low costs of conventionally implanted consumer electronics with 20+ photolithography and implantation steps to make multiple element and (even 3D) concentration gradient structures. The key point: WHY is it necessary to discuss costs in a physics article? There is no need for this, therefore I recommend fully removing this aspect throughout the manuscript, and I give the advice to the authors not to use such an argument at any oral presentation. This again would only polarize the community.

Indeed, we agree and the reference to cost (the phrase "cost effective") was removed. Simply for sake of clarity, what we meant by "cost" is accessibility. I think the referee will agree that presence of FIBs and sputters is by far more common than ion implanters.

(3) Answer letter, page 6, comment to reviewer 2. The sentence "…"no killer applications" have been identified ... and … are limited to effects such as changes in electrical doping of semiconductors" is clear cockiness of the authors on their field. On one side, today there is no single killer application for color centers in diamond, which I can buy in a store, and on the other hand, consumer electronics – including the mobile phones of the authors – rely on the changes in electrical doping of semiconductors by conventional implantation! Again – why do the authors need to polarize the community? There is no need for such a behavior.

Yes, we absolutely agree. That text was only in the response to referee comments, not in the actual paper. The comment about consumer electronics was made by referee 2 ("… computers / mobile phones etc. are today very cheap") – it is not relevant to the paper and we had to respond to it in the letter.

(4) Lines 56 – 58. The sentence "They (… standard implanters …) are also generally ineffective for the implantation of foreign species into atomically-thin materials such as graphene or transition metal dichalcogenides, … " is wrong. The authors should read the reference in detail. True is that any element into any 2D-material can be implanted using standard implanters, because they provide ion energies as low as 20 eV! Up to now there is no single FIB-system, which can provide such low ion energies!

Strictly speaking, we agree with the referee comment and revised this statement. However, the situation isn't as "simple" as the referee portraits. Contamination is still a large challenge (e.g. https://pubs.acs.org/doi/10.1021/acsnano.8b01191). In addition, fit for purpose of commercial ion implantation systems isn't obvious and the community is striving to develop other, rather cumbersome systems (e.g. https://pubs.acs.org/doi/10.1021/acsnano.9b10196)

(5) Even though I appreciate the experiments with nitrogen – supplementray information 2 – these experiments are not convincing at all. Figure S3(b) does not show an PL-intensity increase as a function of ion fluence for the field 1 to 4, and even worse: the PL-yield seems here to be comparable to the not irradiated area between fields 2 and 6, 3 and 7, 4 and 8. Only field 5 shows a clear signal.

This is a good point. The PL spectra do look alike due to normalisation, but the intensity certainly increases. The intensity in between was high due to a non-focused ion beam (scattering). We redid this measurement. The new figure is shown as figure S3.

**References**

[1] Nastasi, M., and J. W. Mayer. "Ion beam mixing in metallic and semiconductor materials." Materials Science and Engineering: R: Reports 12.1 (1994): 1-52.

[2] Lieb, Klaus-Peter et al., "Ion beam mixing in metal/metal and nitride/metal layers-new perspectives." Nuclear Instruments and Methods in Physics Research. Section B, 89.1-4 (1994): 277-289.

[3] Kimura, Tadamasa, et al. "Low-temperature formation of β-type silicon carbide by ion-beam mixing." Japanese journal of applied physics 24.12R (1985): 1712.

[4] Park, Jae-Won et al., "Effects of ion beam mixing of silicon carbide film deposited onto metallic materials for application to nuclear hydrogen production." Journal of nuclear materials 362.2-3 (2007): 268-273.

[5] Khánh, N. Q., et al. "Ion mixing enhanced wafer bonding for silicon‐on‐insulator structures." Journal of applied physics 72.12 (1992): 5602-5605.

[6] Racz, Adél Sarolta, et al. "Novel method for the production of SiC micro and nanopatterns." *Surface and Coatings Technology* 372 (2019): 427-433.

[7] Sadana, D. K., et al. "Shallow doping of gallium arsenide by recoil implantation." *MRS Online Proceedings Library Archive* 147 (1989). Showed doping of GaAs

[8] Liu, Henley L., et al. "Ultra-shallow P+/N junctions formed by recoil implantation." *Journal of electronic materials* 27.9 (1998): 1027-1029. **- NOTE: they actually didn't show a working device**

[9] Christensen, Ove, and Helge L. Bay. "Production of solar cells by recoil implantation." *Applied Physics Letters* 28.9 (1976): 491-494. **- NOTE: they actually didn't show a working device**

[10] Ito, Takashi, et al. "Interfacial Doping by Recoil Implantation for Nonvolatile Memories." *Japanese Journal of Applied Physics* 17.S1 (1978): 201.

[11] Okabayashi, Hidekazu. "Recent progress in refractory-metal silicide formation by ion beam mixing and its applications to VLSL." *Nuclear Instruments and Methods in Physics Research Section B:* 39.1-4 (1989): 246-252.

[12] Feklistov, K. V., et al. "Doping silicon with erbium by recoil implantation." *Technical Physics Letters* 41.8 (2015): 788-792.

[13] Fujimaki, Makoto, et al. "Fabrication of long-period optical fiber gratings by use of ion implantation." *Optics letters* 25.2 (2000): 88-89.

[14] Fujimaki, Makoto, et al. "Ion-implantation-induced densification in silica-based glass for fabrication of optical fiber gratings." *Journal of Applied Physics* 88.10 (2000): 5534-5537.

[15] Yu, Seung Jun, et al. "Birefringence in optical fibers formed by proton implantation." *Nuclear Instruments and Methods in Physics Research Section B:* 265.2 (2007): 490-494.

19th June 2020

Thank you for your letter asking us to reconsider our decision on your manuscript entitled "Knock-on doping: A versatile method to direct-write dopants in a solid state host". Now that I have had a chance to discuss the matter carefully with ___, I am sorry to have to tell you that we do not feel able to reverse our original decision.

Although we do understand the points you raise about some of the statements of reviewer#2, we feel that some of her/his concerns remain valid (at least from our perspective). E.g., there is no doping of the facet of an optical fibre demonstrated, yet there are demonstrations of the preparation of resonant structures and gratings on the facet of optical fibres via FIB in literature [https://doi.org/10.1063/1.3596442, https://www.nature.com/articles/srep15935]. We do not say that this the same, but in our opinion, such demonstrations reduce the nominal advance.

Regarding two other points of the referee you criticise:
from the comments we understood that she/he was pointing at the role of contamination with primary ions from the beam when mentioning uncertainties; And while there are indeed single emitters in your arrays, the implantation is not deterministic but probabilistic at each positions in the array. So we cannot agree that the criticism of reviewer#2 is without substance.

We surely understand that these points can be judged from different perspectives and so can the question of advance.
Therefore, we also understand that different referees come to different conclusions. Yet, let me note that it is in the end an editorial decision, not one made by the referees. We need to gauge the different arguments of the reviewers. And in this case, one of the main reasons why we considered the work was the claim of a new method, which seems to us no longer justified. This is not to say that the presented method will not find application in many labs around the globe. Yet, our judgement remains that this work should be better published in an alternative outlet.

I am sorry that we cannot be more positive on this occasion but hope that you find our referees' comments helpful when preparing your paper for resubmission elsewhere.

24/06/2020
Dear Dr \_\_\_\_,
I've read the final decision sent from \_\_\_. I do understand that the editorial board has their own considerations, **but** those must be based on **facts** and **ethical** merits. In this case, the decision seems to be based on fiction and a clear bias, and I truly sorry we have come to this. I do not ask you to accept our paper, I just ask you to consult with another referee and potentially return it to the 3rd referee for their final opinion.
You have a clear recommendation to publish from ref #1 and only **stylistic** changes from ref #3. Most comments by referee 2 are wrong, biased and misleading and we provided you with objective references. You cannot possibly consider FIB milling of a multimode fibre, as anything even remotely compared to **deterministic** light creation in a core of a single mode fibre – i.e. our work. This is as absurd – like comparing the first transistor (the bulky one!) to any other nanoelectronic device ever made, or comparing any work on diamond NV EPR from the 80s with state of the art NV magnetometry. I trust you know the difference.
While I understand the editorial board "need to gauge the different arguments of the reviewers", the gauging must be based on facts not on fiction! It is obvious referee #2 isn't interested to provide a constructive or an objective advice. There is **NO** contamination in our work (it is proven in our figures!) and the fabricated arrays are deterministic. Like with any method, we are doing a calibration run and repeat it deterministically. If it wasn't deterministic, you wouldn't have had a monotonically increased dose/luminescence signal. The parameters do **NOT change** from run to run. So if you need to base your opinion on a reviewer, it is fine, but please consult an objective one and send it to a 4$^{th}$ referee. Our work is showing something unpredictable with "manipulation and control of materials at truly atomic scales" – just as written in n_xxxxxx aims. I urge you to send the paper to an objective 4$^{th}$ referee and/or reconsider our original appeal. I would also be happy to speak to you over the phone.

22nd July 2020

Dear Igor,

I am finally getting back to you regarding your second letter asking us to revise revert the decision. I have gone through the all the versions and I have thought about it for a while (admittedly it took me longer than anticipated). And I am really sorry but I really do not feel able to change the decision.

I really appreciate the frustration that you may have especially if you disagree with some comments by the reviewers. But eventually we have to look at the overall picture formed by the review process. From the start, when after our conversation we agreed to send it to reviewers our interest was primarily in the practical potential with respect to other techniques. And eventually the overall impression we gather from the reviewers is that such potential is not really justified.